\documentclass[longauth]{aa}

\usepackage{txfonts}
\usepackage{graphicx}
\usepackage{subfigure}
\usepackage{enumerate}
\usepackage{xspace}
\usepackage{makecell}
\usepackage[separate-uncertainty=true,range-phrase=--,range-units=single]{siunitx}
\usepackage[colorlinks,allcolors=blue]{hyperref}
\DeclareSIUnit\pc{pc}
\DeclareSIUnit\kpc{kpc}
\DeclareSIUnit\erg{erg}  
\DeclareSIUnit\year{yr}
\DeclareSIUnit\TeV{TeV}
\DeclareSIUnit\PeV{PeV}
\DeclareSIUnit\gauss{G}
\DeclareSIUnit\Jy{Jy}

\makeatletter
\renewcommand*\aa@pageof{, page \thepage{} of \pageref*{LastPage}}
\makeatother

\begin{document}

\title{A deep spectromorphological study of the $\gamma$-ray emission surrounding the young massive stellar cluster Westerlund~1}
\titlerunning{A deep spectromorphological study of the $\gamma$-ray emission surrounding Westerlund~1}

\author{
\begin{small}
F.~Aharonian \inst{\ref{DIAS},\ref{MPIK}}
\and H.~Ashkar \inst{\ref{LLR}}
\and M.~Backes \inst{\ref{UNAM},\ref{NWU}}
\and V.~Barbosa~Martins \inst{\ref{DESY}}
\and Y.~Becherini \inst{\ref{APC},\ref{Linnaeus}}
\and D.~Berge \inst{\ref{DESY},\ref{HUB}}
\and B.~Bi \inst{\ref{IAAT}}
\and M.~B\"ottcher \inst{\ref{NWU}}
\and M.~de~Bony~de~Lavergne \inst{\ref{LAPP}}
\and F.~Bradascio \inst{\ref{CEA}}
\and R.~Brose \inst{\ref{DIAS}}
\and F.~Brun \inst{\ref{CEA}}
\and T.~Bulik \inst{\ref{UWarsaw}}
\and C.~Burger-Scheidlin \inst{\ref{DIAS}}
\and F.~Cangemi \inst{\ref{LPNHE}}
\and S.~Caroff \inst{\ref{LPNHE}}
\and S.~Casanova \inst{\ref{IFJPAN}}
\and M.~Cerruti \inst{\ref{APC}}
\and T.~Chand \inst{\ref{NWU}}
\and S.~Chandra \inst{\ref{NWU}}
\and A.~Chen \inst{\ref{Wits}}
\and O.~Chibueze \inst{\ref{NWU}}
\and P.~Cristofari \inst{\ref{LUTH}}
\and J.~Damascene~Mbarubucyeye \inst{\ref{DESY}}
\and A.~Djannati-Ata\"i \inst{\ref{APC}}
\and J.-P.~Ernenwein \inst{\ref{CPPM}}
\and K.~Feijen \inst{\ref{Adelaide}}
\and G.~Fichet~de~Clairfontaine \inst{\ref{LUTH}}
\and G.~Fontaine \inst{\ref{LLR}}
\and S.~Funk \inst{\ref{ECAP}}
\and S.~Gabici \inst{\ref{APC}}
\and Y.A.~Gallant \inst{\ref{LUPM}}
\and S.~Ghafourizadeh \inst{\ref{LSW}}
\and G.~Giavitto \inst{\ref{DESY}}
\and L.~Giunti \inst{\ref{APC},\ref{CEA}}
\and D.~Glawion \inst{\ref{ECAP}}
\and J.F.~Glicenstein \inst{\ref{CEA}}
\and P.~Goswami \inst{\ref{NWU}}
\and M.-H.~Grondin \inst{\ref{CENBG}}
\and L.K.~H\"{a}rer \inst{\ref{MPIK}}
\and M.~Haupt \inst{\ref{DESY}}
\and J.A.~Hinton \inst{\ref{MPIK}}
\and M.~H\"{o}rbe \inst{\ref{Oxford}}
\and W.~Hofmann \inst{\ref{MPIK}}
\and T.~L.~Holch \inst{\ref{DESY}}
\and M.~Holler \inst{\ref{Innsbruck}}
\and D.~Horns \inst{\ref{UHH}}
\and M.~Jamrozy \inst{\ref{UJK}}
\and V.~Joshi \inst{\ref{ECAP}}
\and I.~Jung-Richardt \inst{\ref{ECAP}}
\and E.~Kasai \inst{\ref{UNAM}}
\and K.~Katarzy{\'n}ski \inst{\ref{NCUT}}
\and U.~Katz \inst{\ref{ECAP}}
\and B.~Kh\'elifi \inst{\ref{APC}}
\and W.~Klu\'{z}niak \inst{\ref{NCAC}}
\and Nu.~Komin \inst{\ref{Wits}}
\and K.~Kosack \inst{\ref{CEA}}
\and D.~Kostunin \inst{\ref{DESY}}
\and G.~Kukec~Mezek \inst{\ref{Linnaeus}}
\and R.G.~Lang \inst{\ref{ECAP}}
\and S.~Le Stum \inst{\ref{CPPM}}
\and A.~Lemi\`ere \inst{\ref{APC}}
\and M.~Lemoine-Goumard \inst{\ref{CENBG}}
\and J.-P.~Lenain \inst{\ref{LPNHE}}
\and F.~Leuschner \inst{\ref{IAAT}}
\and T.~Lohse \inst{\ref{HUB}}
\and A.~Luashvili \inst{\ref{LUTH}}
\and I.~Lypova \inst{\ref{LSW}}
\and J.~Mackey \inst{\ref{DIAS}}
\and J.~Majumdar \inst{\ref{DESY}}
\and D.~Malyshev \inst{\ref{ECAP}}
\and V.~Marandon \inst{\ref{MPIK}}
\and P.~Marchegiani \inst{\ref{Wits}}
\and A.~Marcowith \inst{\ref{LUPM}}
\and G.~Mart\'i-Devesa \inst{\ref{Innsbruck}}
\and R.~Marx \inst{\ref{LSW}}
\and G.~Maurin \inst{\ref{LAPP}}
\and M.~Meyer \inst{\ref{UHH}}
\and A.~Mitchell \inst{\ref{ECAP},\ref{MPIK}}
\and R.~Moderski \inst{\ref{NCAC}}
\and L.~Mohrmann \inst{\ref{MPIK}}$^{,}$\thanks{Corresponding authors;\\\email{\href{mailto:contact.hess@hess-experiment.eu}{contact.hess@hess-experiment.eu}}}
\and A.~Montanari \inst{\ref{CEA}}
\and E.~Moulin \inst{\ref{CEA}}
\and J.~Muller \inst{\ref{LLR}}
\and T.~Murach \inst{\ref{DESY}}
\and K.~Nakashima \inst{\ref{ECAP}}
\and M.~de~Naurois \inst{\ref{LLR}}
\and A.~Nayerhoda \inst{\ref{IFJPAN}}
\and J.~Niemiec \inst{\ref{IFJPAN}}
\and S.~Ohm \inst{\ref{DESY}}$^{,}$\footnotemark[1] % Corresponding author
\and L.~Olivera-Nieto \inst{\ref{MPIK}}
\and E.~de~Ona~Wilhelmi \inst{\ref{DESY}}
\and M.~Ostrowski \inst{\ref{UJK}}
\and S.~Panny \inst{\ref{Innsbruck}}
\and M.~Panter \inst{\ref{MPIK}}
\and R.D.~Parsons \inst{\ref{HUB}}
\and G.~Peron \inst{\ref{MPIK}}
\and D.A.~Prokhorov \inst{\ref{Amsterdam}}
\and G.~P\"uhlhofer \inst{\ref{IAAT}}
\and M.~Punch \inst{\ref{APC},\ref{Linnaeus}}
\and A.~Quirrenbach \inst{\ref{LSW}}
\and R.~Rauth \inst{\ref{Innsbruck}}$^{,}$\footnotemark[1] % Corresponding author
\and P.~Reichherzer \inst{\ref{CEA}}
\and A.~Reimer \inst{\ref{Innsbruck}}
\and O.~Reimer \inst{\ref{Innsbruck}}
\and M.~Renaud \inst{\ref{LUPM}}
\and B.~Reville \inst{\ref{MPIK}}
\and F.~Rieger \inst{\ref{MPIK}}
\and G.~Rowell \inst{\ref{Adelaide}}
\and B.~Rudak \inst{\ref{NCAC}}
\and E.~Ruiz-Velasco \inst{\ref{MPIK}}
\and V.~Sahakian \inst{\ref{Yerevan}}
\and H.~Salzmann \inst{\ref{IAAT}}
\and D.A.~Sanchez \inst{\ref{LAPP}}
\and A.~Santangelo \inst{\ref{IAAT}}
\and M.~Sasaki \inst{\ref{ECAP}}
\and F.~Sch\"ussler \inst{\ref{CEA}}
\and H.M.~Schutte \inst{\ref{NWU}}
\and U.~Schwanke \inst{\ref{HUB}}
\and J.N.S.~Shapopi \inst{\ref{UNAM}}
\and A.~Specovius \inst{\ref{ECAP}}$^{,}$\footnotemark[1] % Corresponding author
\and S.~Spencer \inst{\ref{Oxford}}
\and {\L.}~Stawarz \inst{\ref{UJK}}
\and R.~Steenkamp \inst{\ref{UNAM}}
\and S.~Steinmassl \inst{\ref{MPIK}}
\and C.~Steppa \inst{\ref{UP}}
\and I.~Sushch \inst{\ref{NWU}}
\and H.~Suzuki \inst{\ref{Konan}}
\and T.~Takahashi \inst{\ref{KAVLI}}
\and T.~Tanaka \inst{\ref{Konan}}
\and R.~Terrier \inst{\ref{APC}}
\and C.~Thorpe-Morgan \inst{\ref{IAAT}}
\and M.~Tsirou \inst{\ref{MPIK}}
\and N.~Tsuji \inst{\ref{RIKKEN}}
\and R.~Tuffs \inst{\ref{MPIK}}
\and T.~Unbehaun \inst{\ref{ECAP}}
\and C.~van~Eldik \inst{\ref{ECAP}}
\and B.~van~Soelen \inst{\ref{UFS}}
\and M.~Vecchi \inst{\ref{Groningen}}
\and J.~Veh \inst{\ref{ECAP}}
\and C.~Venter \inst{\ref{NWU}}
\and J.~Vink \inst{\ref{Amsterdam}}
\and S.J.~Wagner \inst{\ref{LSW}}
\and R.~White \inst{\ref{MPIK}}
\and A.~Wierzcholska \inst{\ref{IFJPAN}}
\and Yu~Wun~Wong \inst{\ref{ECAP}}
\and M.~Zacharias \inst{\ref{LUTH},\ref{NWU}}
\and D.~Zargaryan \inst{\ref{DIAS}}
\and A.A.~Zdziarski \inst{\ref{NCAC}}
\and S.J.~Zhu \inst{\ref{DESY}}
\and S.~Zouari \inst{\ref{APC}}
\and N.~\.Zywucka \inst{\ref{NWU}} (H.E.S.S.\ Collaboration)
\and \\R.~Blackwell \inst{\ref{Adelaide}}
\and C.~Braiding \inst{\ref{Sydney}}
\and M.~Burton \inst{\ref{Armagh}}
\and K.~Cubuk \inst{\ref{Armagh}}
\and M.~Filipovi\'{c} \inst{\ref{WesternSydney}}
\and N.~Tothill \inst{\ref{WesternSydney}}
\and G.~Wong \inst{\ref{WesternSydney}}
\end{small}
}
\authorrunning{F. Aharonian et al.}

\institute{
Dublin Institute for Advanced Studies, 31 Fitzwilliam Place, Dublin 2, Ireland \label{DIAS} \and
Max-Planck-Institut f\"ur Kernphysik, P.O. Box 103980, D 69029 Heidelberg, Germany \label{MPIK} \and
Laboratoire Leprince-Ringuet, École Polytechnique, CNRS, Institut Polytechnique de Paris, F-91128 Palaiseau, France \label{LLR} \and
University of Namibia, Department of Physics, Private Bag 13301, Windhoek 10005, Namibia \label{UNAM} \and
Centre for Space Research, North-West University, Potchefstroom 2520, South Africa \label{NWU} \and
DESY, D-15738 Zeuthen, Germany \label{DESY} \and
Université de Paris, CNRS, Astroparticule et Cosmologie, F-75013 Paris, France \label{APC} \and
Department of Physics and Electrical Engineering, Linnaeus University,  351 95 V\"axj\"o, Sweden \label{Linnaeus} \and
Institut f\"ur Physik, Humboldt-Universit\"at zu Berlin, Newtonstr. 15, D 12489 Berlin, Germany \label{HUB} \and
Institut f\"ur Astronomie und Astrophysik, Universit\"at T\"ubingen, Sand 1, D 72076 T\"ubingen, Germany \label{IAAT} \and
Université Savoie Mont Blanc, CNRS, Laboratoire d'Annecy de Physique des Particules - IN2P3, 74000 Annecy, France \label{LAPP} \and
IRFU, CEA, Universit\'e Paris-Saclay, F-91191 Gif-sur-Yvette, France \label{CEA} \and
Astronomical Observatory, The University of Warsaw, Al. Ujazdowskie 4, 00-478 Warsaw, Poland \label{UWarsaw} \and
Sorbonne Universit\'e, Universit\'e Paris Diderot, Sorbonne Paris Cit\'e, CNRS/IN2P3, Laboratoire de Physique Nucl\'eaire et de Hautes Energies, LPNHE, 4 Place Jussieu, F-75252 Paris, France \label{LPNHE} \and
Instytut Fizyki J\c{a}drowej PAN, ul. Radzikowskiego 152, 31-342 Krak{\'o}w, Poland \label{IFJPAN} \and
School of Physics, University of the Witwatersrand, 1 Jan Smuts Avenue, Braamfontein, Johannesburg, 2050 South Africa \label{Wits} \and
Laboratoire Univers et Théories, Observatoire de Paris, Université PSL, CNRS, Université de Paris, 92190 Meudon, France \label{LUTH} \and
Aix Marseille Universit\'e, CNRS/IN2P3, CPPM, Marseille, France \label{CPPM} \and
School of Physical Sciences, University of Adelaide, Adelaide 5005, Australia \label{Adelaide} \and
Friedrich-Alexander-Universit\"at Erlangen-N\"urnberg, Erlangen Centre for Astroparticle Physics, Erwin-Rommel-Str. 1, D 91058 Erlangen, Germany \label{ECAP} \and
Laboratoire Univers et Particules de Montpellier, Universit\'e Montpellier, CNRS/IN2P3,  CC 72, Place Eug\`ene Bataillon, F-34095 Montpellier Cedex 5, France \label{LUPM} \and
Landessternwarte, Universit\"at Heidelberg, K\"onigstuhl, D 69117 Heidelberg, Germany \label{LSW} \and
Universit\'e Bordeaux, CNRS, LP2I Bordeaux, UMR 5797, F-33170 Gradignan, France \label{CENBG} \and
University of Oxford, Department of Physics, Denys Wilkinson Building, Keble Road, Oxford OX1 3RH, UK \label{Oxford} \and
Institut f\"ur Astro- und Teilchenphysik, Leopold-Franzens-Universit\"at Innsbruck, A-6020 Innsbruck, Austria \label{Innsbruck} \and
Universit\"at Hamburg, Institut f\"ur Experimentalphysik, Luruper Chaussee 149, D 22761 Hamburg, Germany \label{UHH} \and
Obserwatorium Astronomiczne, Uniwersytet Jagiello{\'n}ski, ul. Orla 171, 30-244 Krak{\'o}w, Poland \label{UJK} \and
Institute of Astronomy, Faculty of Physics, Astronomy and Informatics, Nicolaus Copernicus University,  Grudziadzka 5, 87-100 Torun, Poland \label{NCUT} \and
Nicolaus Copernicus Astronomical Center, Polish Academy of Sciences, ul. Bartycka 18, 00-716 Warsaw, Poland \label{NCAC} \and
GRAPPA, Anton Pannekoek Institute for Astronomy, University of Amsterdam,  Science Park 904, 1098 XH Amsterdam, The Netherlands \label{Amsterdam} \and
Yerevan Physics Institute, 2 Alikhanian Brothers St., 375036 Yerevan, Armenia \label{Yerevan} \and
Institut f\"ur Physik und Astronomie, Universit\"at Potsdam,  Karl-Liebknecht-Strasse 24/25, D 14476 Potsdam, Germany \label{UP} \and
Department of Physics, Konan University, 8-9-1 Okamoto, Higashinada, Kobe, Hyogo 658-8501, Japan \label{Konan} \and
Kavli Institute for the Physics and Mathematics of the Universe (WPI), The University of Tokyo Institutes for Advanced Study (UTIAS), The University of Tokyo, 5-1-5 Kashiwa-no-Ha, Kashiwa, Chiba, 277-8583, Japan \label{KAVLI} \and
RIKEN, 2-1 Hirosawa, Wako, Saitama 351-0198, Japan \label{RIKKEN} \and
Department of Physics, University of the Free State,  PO Box 339, Bloemfontein 9300, South Africa \label{UFS} \and
Kapteyn Astronomical Institute, University of Groningen, Landleven 12, 9747 AD Groningen, The Netherlands \label{Groningen} \and
School of Physics, University of New South Wales, Sydney, NSW 2052, Australia \label{Sydney} \and
Armagh Observatory and Planetarium, Armagh, BT61 7HT, Northern Ireland, United Kingdom \label{Armagh} \and
Western Sydney University, Locked Bag 1797 Penrith South DC, NSW 2751, Australia \label{WesternSydney}
}

\date{\today}

\abstract
{% Context
  Young massive stellar clusters are extreme environments and potentially provide the means for efficient particle acceleration.
  Indeed, they are increasingly considered as being responsible for a significant fraction of cosmic rays (CRs) accelerated within the Milky Way.
  Westerlund~1, the most massive known young stellar cluster in our Galaxy is a prime candidate for studying this hypothesis.
  While the very-high-energy $\gamma$-ray source HESS~J1646$-$458 has been detected in the vicinity of Westerlund~1 in the past, its association could not be firmly identified.
}
{% Aims
  We aim to identify the physical processes responsible for the $\gamma$-ray emission around Westerlund~1 and thus to better understand the role of massive stellar clusters in the acceleration of Galactic CRs.
}
{% Methods
  Using 164~hours of data recorded with the High Energy Stereoscopic System (H.E.S.S.), we carried out a deep spectromorphological study of the $\gamma$-ray emission of HESS~J1646$-$458.
  We furthermore employed H~I and CO observations of the region to infer the presence of gas that could serve as target material for interactions of accelerated CRs.
}
{% Results
  We detected large-scale ($\sim$$2^\circ$ diameter) $\gamma$-ray emission with a complex morphology, exhibiting a shell-like structure and showing no significant variation with $\gamma$-ray energy.
  The combined energy spectrum of the emission extends to several tens of TeV, and is uniform across the entire source region.
  We did not find a clear correlation of the $\gamma$-ray emission with gas clouds as identified through H~I and CO observations.
}
{% Conclusions
  We conclude that, of the known objects within the region, only Westerlund~1 can explain the bulk of the $\gamma$-ray emission.
  Several CR acceleration sites and mechanisms are conceivable, and discussed in detail.
  While it seems clear that Westerlund~1 acts as a powerful particle accelerator, no firm conclusions on the contribution of massive stellar clusters to the flux of Galactic CRs in general can be drawn at this point.
}


\maketitle

\section{Introduction}

Young massive stellar clusters are environments of copious star formation, and typically host a large number of very massive stars \citep{PortegiesZwart2010}.
For this reason, they have long been considered as potential sites of cosmic-ray (CR) acceleration \citep{Parizot2004}.
The acceleration may take place at shock fronts of supernova remnants (SNRs) \citep[which may collide with the strong winds of massive stars in the cluster, see e.g.\xspace][and references therein]{Bykov2020}, or at the termination shock of the superbubble that is excavated by the combined stellar winds of the cluster \citep[e.g.][]{Bykov2014,Gupta2018,Morlino2021}.
Massive clusters form from correspondingly massive molecular clouds, which are not very common in the Milky Way but often found in starburst galaxies \citep{Fujii2016}.
Nevertheless, the notion that massive star clusters are responsible for the bulk of hadronic CRs\footnote{
  Here and in the following, the term ``hadronic cosmic ray'' refers to cosmic-ray nuclei, as opposed to, e.g., cosmic-ray electrons and positrons.
}
accelerated within our Galaxy represents a viable alternative to the long-standing ``SNR paradigm'', in which (isolated) SNRs are the primary acceleration sites \citep{PortegiesZwart2010,Aharonian2019,Morlino2021}.

Through interaction with ambient gas and radiation fields, high-energy hadronic CRs produce high-energy $\gamma$ rays, which provides strong motivation for the search for $\gamma$-ray emission from massive stellar clusters \citep[this has been realised already long ago, see e.g.][]{Cesarsky1983}.
Indeed, a bubble, or ``cocoon'' in the Cygnus~X star-forming region has been detected in \emph{Fermi}-LAT data in the $\sim$\SI{1}{\GeV}--few~\SI{100}{\GeV} energy range \citep{FermiLAT2011}.
Subsequently, \emph{Fermi}-LAT has detected diffuse $\gamma$-ray emission in the same energy range around a number of other massive stellar clusters in the Milky Way \citep{Yang2017,Yang2018,Yang2020,Sun2020,Sun2020a}.
Searches at higher energies (i.e.\ at \SI{1}{\TeV} and above) with ground-based instruments have also lead to several detections:
\begin{itemize}
  \item the Cygnus region harbours several sources of TeV-energy $\gamma$ rays \citep{Abdo2007,VERITAS2018}, and has recently been detected up to energies of hundreds of TeV \citep{HAWC2021,LHAASO2021};
  \item the young stellar cluster Westerlund~2 within the star formation region RCW~49, which hosts with WR~20a an extraordinarily massive binary star system \citep{HESS_Wd2_2007,HESS_Wd2_2011};
  \item the young stellar cluster Westerlund~1 \citep{HESS_Westerlund1_2012}, which will be introduced in more detail below;
  \item the super bubble 30 Dor C, whose detection is particularly noteworthy due to its distant location in the Large Magellanic Cloud \citep{HESS_30DorC_2015};
  \item the stellar cluster Cl$^\ast$~1806$-$20, which contains both a luminous blue variable candidate, LBV~1806$-$20, and a magnetar, SGR~1806$-$20 \citep{HESS_MassiveStars_2018}.
\end{itemize}
However, the $\gamma$-ray emission can be linked directly to the stellar clusters in only some of the above cases, the precise CR acceleration sites are yet unidentified, and none of the detections constitutes unequivocal evidence for the acceleration of hadronic CRs due to the respective clusters.
The assertion of the latter point is complicated by the fact that high-energy $\gamma$ rays can be produced by CRs via two competing processes.
Besides their production in the decay of neutral pions (and other short-lived particles), produced in turn in interactions of hadronic CRs with ambient matter -- the ``hadronic scenario'' -- they may also be created through the inverse Compton (IC) process, in which high-energy electrons and positrons\footnote{
  Hereafter, we use the term ``electrons'' to refer to both electrons and positrons.
}
can up-scatter low-energy photons from ambient radiation fields to TeV energies -- the ``leptonic scenario''.
These two scenarios can only be distinguished by carrying out detailed spectromorphological studies of the $\gamma$-ray emission, and combining the results with those obtained at other wavelengths.
In this article, we present such a study for the young massive stellar cluster Westerlund~1.

Westerlund~1, named after its discoverer Bengt Westerlund \citep{Westerlund1961} and located at R.A.(2000)=$16^\mathrm{h}47^\mathrm{m}04.0^\mathrm{s}$, Dec.(2000)=$-45^\circ 51'04.9''$ \citep{Brandner2008}, is the most massive known young stellar cluster in the Milky Way, with an estimated mass of around $10^5\,M_\odot$ \citep{Clark2005,Brandner2008,PortegiesZwart2010}.
It hosts a rich population of evolved massive stars, including significant fractions of all known Galactic Yellow Hypergiants \citep{Clark2005} and Wolf-Rayet stars \citep{Crowther2006}.
The half-mass radius of the cluster is approximately \SI{1}{\pc} \citep{Brandner2008}.
Many estimates for the age of the cluster and its distance from Earth have been put forward in the past.
Most age estimates agree with an age of \SIrange{3}{5}{\mega\year} \citep{Clark2005,Crowther2006,Brandner2008}, although the single-age paradigm has been questioned recently after finding that the observed luminosities of cool supergiants are more consistent with an age of $\sim$\SI{10}{\mega\year} \citep{Beasor2021}.
Early distance estimates, using various techniques, find distances of around \SI{4}{\kpc} \citep{Clark2005,Crowther2006,Kothes2007,Brandner2008}.
Recently, data from the \emph{Gaia} spacecraft were used to obtain new distance estimates.
While most of them are compatible with the old estimates \citep{Davies2019,Rate2020,Beasor2021,Negueruela2022}, closer distances of $\sim$\SI{2.7}{\kpc} have also been obtained \citep{Aghakhanloo2020,Aghakhanloo2021}.
\citet{Clark2019} have questioned the reliability of \emph{Gaia} (DR2) data in the Westerlund~1 field altogether, rendering the new estimates somewhat uncertain.
For this article, we adopted an age of \SI{4}{\mega\year} and a distance of \SI{3.9}{\kpc}, as these values are compatible with the majority of published results.
Additionally, we will need in the course of this paper estimates for the properties of the collective cluster wind of Westerlund~1, which is formed as a superposition of the strong winds of the massive stars in the cluster.
Only few estimates can be found in the literature; we assumed here for the kinetic luminosity of the wind $L_\mathrm{w}\sim$\SI{e39}{\erg\per\second} \citep{Muno2006} and for the wind velocity $v_\mathrm{w}\sim$\SI{3000}{\km\per\second} \citep{Morlino2021} as typical values.
All parameter values for Westerlund~1 assumed in this work are summarised in Table~\ref{tab:wd1_pars}.

\begin{table}
  \centering
  \caption{Parameter values for Westerlund~1 assumed in this work.}
  \label{tab:wd1_pars}
  \begin{tabular}{cccc}
    \hline\hline
    Par. & Description & Value & Ref.\\\hline
    $d$ & distance from Earth & \SI{3.9}{\kpc} & (1,2) \\
    $\tau$ & cluster age & \SI{4}{\mega\year} & (3,4) \\
    $L_\mathrm{w}$ & \makecell{kinetic luminosity\\of cluster wind} & \SI{e39}{\erg\per\second} & (5) \\
    $v_\mathrm{w}$ & velocity of cluster wind & \SI{3000}{\km\per\second} & (6) \\
    \hline
  \end{tabular}
  \tablebib{
    (1) \citet{Kothes2007}; (2) \citet{Davies2019}; (3) \citet{Clark2005};
    (4) \citet{Brandner2008}; (5) \citet{Muno2006}; (6) \citet{Morlino2021}.
   }
\end{table}

Westerlund~1 has been studied extensively in the X-ray domain.
Observations with the \emph{Chandra} telescope have revealed diffuse hard X-ray emission from the core of the cluster \citep{Muno2006}, which was later identified as likely being of thermal origin with \emph{XMM-Newton} observations \citep{Kavanagh2011}.
Additionally, \citet{Muno2006a} have identified an X-ray magnetar, CXOU~J164710.2$-$455216, which supposedly was created in the explosion of a very massive ($>40\,M_\odot$) progenitor star \citep{Clark2008,Belczynski2008}.
Interestingly, CXOU~J164710.2$-$455216 is the only known remnant of a stellar explosion within Westerlund~1.

Moving to larger spatial scales (i.e.\ beyond the bounds of the cluster itself), an analysis of \emph{Fermi}-LAT data between 3 and \SI{300}{\GeV} by \citet{Ohm2013} revealed extended $\gamma$-ray emission in the vicinity of Westerlund~1.
With more data accumulated since then, the latest \emph{Fermi}-LAT source catalogue \citep[4FGL-DR2,][]{FermiLAT2020,FermiLAT2020a} lists six sources within $1.1^\circ$ from the cluster centre.
Besides the stellar cluster and its members, several objects that are located at relatively small angular separations from Westerlund~1 could potentially be contributing to the observed $\gamma$-ray emission.
This includes two energetic ($\dot{E}>\SI{2e35}{\erg\per\second}$) pulsars, PSR~J1648$-$4611 and PSR~J1650$-$4601 \citep{Manchester2005}, as well as the low-mass X-ray binary (LMXB) 4U~1642$-$45 \citep{Forman1978}.
On the other hand, it is quite possible that some of the six sources listed in the 4FGL catalogue share a common physical origin, and were separated into distinct components only because the true source morphology is very complex.

Finally, at even higher energies, \citet{HESS_Westerlund1_2012} detected a large, extended ($\sim$$2^\circ$ diameter) emission region centred on Westerlund~1, named HESS~J1646$-$458, between 0.45 and $\sim$\SI{20}{\TeV} with the H.E.S.S.\ experiment.
Based on the properties of the emission and taking into account multi-wavelength data, the authors found Westerlund~1 to be the most likely explanation of the $\gamma$-ray emission in a single-source scenario, but were unable to draw definitive conclusions based on the data set available at the time.

Since then, the exposure collected with H.E.S.S.\ on HESS~J1646$-$458 has almost quintupled, in large part thanks to a dedicated observation campaign in 2017.
This, together with recent advances in analysis techniques, enabled a new, detailed study of the $\gamma$-ray emission surrounding Westerlund~1, which we present here.
In Sect.~\ref{sec:hess_data_analysis}, we introduce the H.E.S.S.\ data set and provide a description of the data analysis.
The results of the H.E.S.S.\ data analysis are given in Sect.~\ref{sec:hess_results}.
Besides H.E.S.S.\ data, we have also analysed data from H~I and CO observations in the vicinity of Westerlund~1 and present the results in Sect.~\ref{sec:radio}.
A detailed discussion of the results, considering multiple explanations for the observed $\gamma$-ray emission, is presented in Sect.~\ref{sec:discussion}, before we summarise our findings and provide an outlook in Sect.~\ref{sec:conclusion}.

\section{Observations and data analysis}

\subsection{H.E.S.S.\ data set and analysis}
\label{sec:hess_data_analysis}

H.E.S.S.\ is an array of five imaging atmospheric Cherenkov telescopes (IACTs), located in the Southern hemisphere in Namibia ($23^\circ 16'18''$~S, $16^\circ 30'00''$~E) at an altitude of \SI{1800}{\meter} above sea level \citep{HESS_Crab_2006,Holler2015}.
The array comprises four \SI{12}{\meter}-diameter telescopes (CT1-4) that are arranged on a \SI{120}{\meter}-side square and began operation in late 2003.
A fifth telescope (CT5), with \SI{28}{\meter} diameter, was added in the centre of the array in 2012.
The telescopes detect the Cherenkov light produced in atmospheric air showers initiated by primary $\gamma$ rays, where the main background consists of showers caused by hadronic CR.
With the central telescope included, the array is sensitive to $\gamma$ rays in the energy range between $\sim$\SI{0.1}{\TeV} and $\sim$\SI{100}{\TeV}; with CT5 alone thresholds as low as \SI{20}{\GeV} have been achieved in studies of pulsed emission \citep{HESS_VelaPulsar_2018}.

The H.E.S.S.\ data set for HESS~J1646$-$458 comprises 362~observation runs after quality selection, taken over the course of more than 13~years between June~18, 2004 and October~11, 2017.
The runs amount to a total observation time of 164.2~hours, which represents a significant increase with respect to the previous publication \citep{HESS_Westerlund1_2012} (33.8~hours).
We note that not all of the observations have targeted HESS~J1646$-$458 directly; some have been taken as part of surveys, and some were primarily targeted at the nearby sources HESS~J1640$-$465 \citep{HESS_1640_2014} and HESS~J1641$-$463 \citep{HESS_1641_2014}, leading to a gradient in exposure across the HESS~J1646$-$458 region.

Only data from the four smaller telescopes (CT1-4) were considered in the analysis presented here.
For best performance, we restricted the maximum zenith angle of the analysed observation runs to $<60^\circ$ and the maximum offset angle between the reconstructed direction of events and the telescope pointing direction to $<2^\circ$.
With this selection, an energy threshold of \SI{0.37}{\TeV} was achieved in the final analysis.
$\gamma$-like events were selected using the method described in \citet{Ohm2009} and their energy and arrival direction were reconstructed with the algorithm presented in \citet{Parsons2014}.
Subsequently, we converted our data to the FITS-based data format described in \citet{Deil2018} and performed the high-level data analysis using the \textsc{Gammapy} package \citep{Deil2017,Deil2020} (v0.17).
All findings were confirmed with two cross-check analyses: one based on a completely independent calibration and data analysis chain \citep{deNaurois2009}, and one based on the same calibration and event reconstruction algorithms as those used in the main analysis, but carried out with the \textsc{ctools} package \citep{Knoedlseder2016} (v.1.6.3); the latter analysis is documented in \citet{Specovius2021}.
Furthermore, results of another, intermediate analysis of the data set, which has inspired parts of the analysis presented here, can be found in \citet{Zorn2019}.

In the high-level analysis, we employed a concept that has only recently been established for the analysis of IACT data: a 3-dimensional likelihood analysis, in which the data can be modelled simultaneously in two spatial dimensions and as a function of energy \citep{Mohrmann2019}.
In this method, contrary to more established ones, the residual background of CR-induced air shower events (``hadronic'' background) for a given observation run is not directly estimated from source-free regions within the observed field itself, but rather provided by a background model.
This model was constructed from archival observations and subsequently adjusted to the analysed observations following the procedure outlined in \citet{Mohrmann2019}.
The 3-dimensional likelihood analysis method is especially suited for the analysis of complex source regions and largely extended sources, and is thus a suitable choice for the analysis of HESS~J1646$-$458.

As a first step in the analysis, separate energy thresholds were determined for each observation run, requiring that the energy reconstruction bias is below 10\% and that the background model is used only above its validity threshold \citep{Mohrmann2019}.
Aiming for sufficient exposure across the entire region down to the lowest energies, a minimal energy threshold of \SI{0.37}{\TeV} was enforced.
We subsequently computed 3-dimensional maps of the observed number of events, predicted background, and exposure, comprised of spatial pixels of $0.02^\circ \times 0.02^\circ$ and an energy axis of 16~bins per decade of energy, with the $6^\circ \times 6^\circ$ region of interest centred on Westerlund~1.
For each observation, we adjusted the background model to the observed data by fitting a global normalisation and a spectral tilt parameter, taking into account only regions in the field of view that are free of $\gamma$-ray emission.
In order to safely exclude regions with $\gamma$-ray emission from the background fit, an iterative procedure as described in \citet{HESS_HGPS_2018} was employed to generate an exclusion map.

\subsection{H.E.S.S.\ analysis results}
\label{sec:hess_results}

\subsubsection{Maps and radial profiles}
\label{sec:maps_profiles}

\begin{figure*}[ht]
  \centering
  \subfigure[]{
    \includegraphics{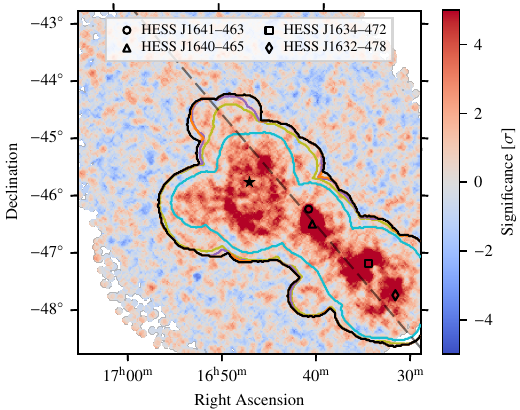}
    \label{fig:sign_map_div}
  }
  \subfigure[]{
    \includegraphics{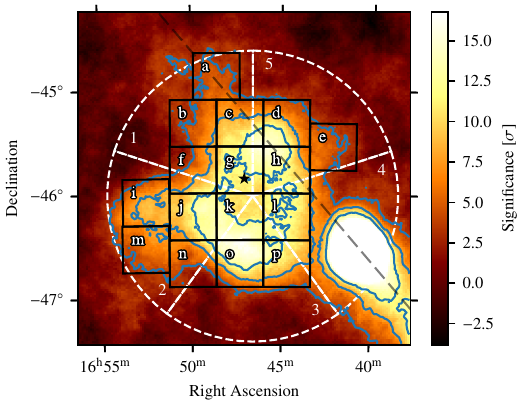}
    \label{fig:sign_map_boxes}
  }
  \caption{
    Significance maps after background subtraction.
    The position of Westerlund~1 is marked by the black star symbol; the grey, dashed line shows the Galactic plane.
    (a) Map for the entire $6^\circ \times 6^\circ$ region of interest, smoothed with a $0.07^\circ$ top-hat kernel.
    The final exclusion map is shown in black, earlier iterations in blue, green, purple, and orange.
    Locations of previously detected sources that are not connected to HESS~J1646$-$458 are indicated by black, open symbols.
    (b) Map with detail view of the emission surrounding Westerlund~1, smoothed with a $0.22^\circ$ top-hat kernel.
    The colour scale is saturated at the maximum observed significance value associated with the HESS~J1646$-$458 region.
    Contour lines corresponding to a significance of 4, 8, and 12 $\sigma$ are shown in blue.
    Signal regions a--p used for spectrum extraction are overlaid (black), as are segments 1--5 for the computation of radial profiles (white dashed).
  }
  \label{fig:sign_maps}
\end{figure*}

\begin{figure}
  \centering
  \includegraphics{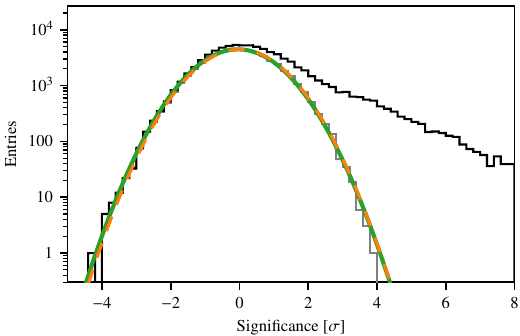}
  \caption{
    Significance entry distribution.
    The black histogram corresponds to all pixels of the map shown in Fig.~\ref{fig:sign_map_div} with non-zero entries, the grey histogram to all pixels outside of the final exclusion map.
    The green line represents the fit of a Gaussian distribution to the grey histogram, the best-fit mean and width are $\mu=-0.043\pm 0.005$ and $\sigma=1.008\pm 0.005$, respectively.
    A Gaussian distribution with mean $\mu=0$ and width $\sigma=1$ is shown by the orange, dashed line for comparison.
  }
  \label{fig:sign_dist}
\end{figure}

We show in Fig.~\ref{fig:sign_maps} the resulting residual significance maps after background subtraction, where the significance was computed following \citet{Li1983}.
For the map in Fig.~\ref{fig:sign_map_div}, a top-hat smoothing with a kernel of radius $0.07^\circ$ has been employed.
The corresponding distribution of significance values -- for all pixels and those outside the exclusion map, displayed in black -- is presented in Fig.~\ref{fig:sign_dist}.
As expected for the case of purely statistical fluctuations, the distribution for pixels outside the exclusion map follows very closely that of a Gaussian distribution with unit width, indicating a good description of the hadronic background.
Figure~\ref{fig:sign_map_boxes} shows a significance map obtained with a correlation radius of $0.22^\circ$.
Overlaid are 16~square ``signal regions'' (of size $0.45^\circ\times 0.45^\circ$ each), labelled a--p, that cover the $\gamma$-ray emission of HESS~J1646$-$458, as well as 5~circular segments.
These signal regions and segments are used in the further characterisation of the $\gamma$-ray emission (see below).

\begin{figure*}
  \centering
  \subfigure[Smoothing kernel: $0.22^\circ$ top hat]{
    \includegraphics{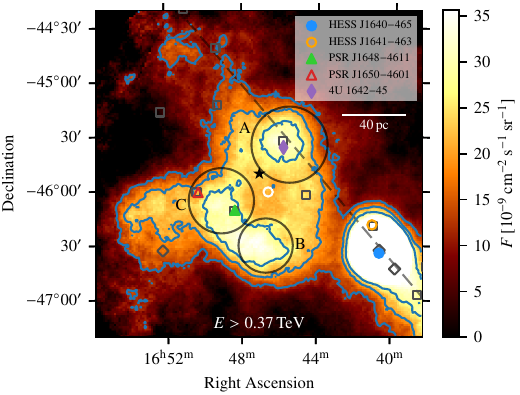}
    \label{fig:flux_map}
  }
  \subfigure[Smoothing kernel: $0.07^\circ$ Gaussian]{
    \includegraphics{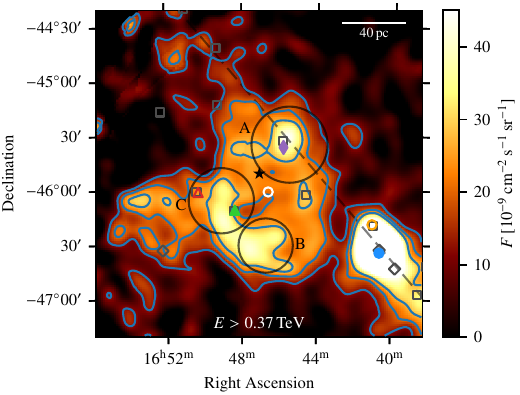}
    \label{fig:flux_map_hires}
  }\\
  \subfigure[Smoothing kernel: $0.22^\circ$ top hat]{
    \includegraphics{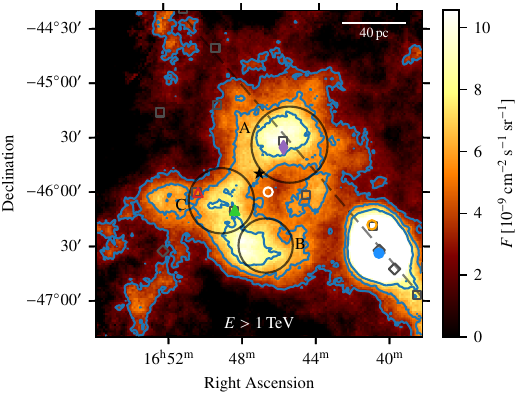}
    \label{fig:flux_map_1TeV}
  }
  \subfigure[Smoothing kernel: $0.22^\circ$ top hat]{
    \includegraphics{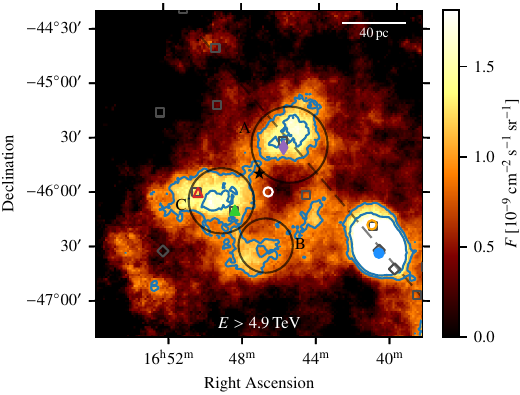}
    \label{fig:flux_map_5TeV}
  }
  \caption{
    Flux maps of the HESS~J1646$-$458 region.
    The position of Westerlund~1 is marked by the black star symbol; the grey, dashed line shows the Galactic plane.
    Coloured symbols indicate objects listed in the legend in panel (a).
    Dark grey square markers denote positions of sources from the 4FGL-DR2 catalogue \citep{FermiLAT2020,FermiLAT2020a}, where those sources that are still significant ($\sqrt{\mathrm{TS}}>3$) above \SI{30}{\GeV} are shown with a diamond marker ($\Diamond$).
    Grey circles labelled `A' and `B' mark regions defined in \citet{HESS_Westerlund1_2012}; region `C' (at R.A.~$16^\mathrm{h}49^\mathrm{m}4.8^\mathrm{s}$, Dec.~$-46^\circ 06'00''$) is newly defined here.
    The white circle marker indicates the coordinate with respect to which the radial profiles in Fig.~\ref{fig:radial_profiles} and \ref{fig:cr_dens_profile_shifted} have been computed.
    The scale bar denotes a projected distance of \SI{40}{\pc}, for the nominal distance to Westerlund~1 of \SI{3.9}{\kpc}.
    The maps are for different energy thresholds (indicated at the bottom of each panel) and were computed using different smoothing kernels (stated below each figure).
    Colour scales are saturated at the maximum observed flux value associated with the HESS~J1646$-$458 region.
    Contour lines shown in blue are at flux levels of $F=(12.5 / 20 / 27.5)\times 10^{-9}\si{\per\square\cm\per\second\per\steradian}$ for panels (a) and (b), at $F=(3 / 5.5 / 8)\times 10^{-9}\si{\per\square\cm\per\second\per\steradian}$ for panel (c), and at $F=(1 / 1.5)\times 10^{-9}\si{\per\square\cm\per\second\per\steradian}$ for panel (d).
  }
  \label{fig:flux_maps}
\end{figure*}

Having obtained a satisfying description of the residual hadronic background, we computed flux maps displaying the excess $\gamma$-ray emission (see Fig.~\ref{fig:flux_maps}).
Focusing on the larger-scale structure of the emission of HESS~J1646$-$458, a top-hat smoothing with a kernel of radius~$0.22^\circ$ -- the same value as already used in \citet{HESS_Westerlund1_2012} -- has been applied for the maps in panels~\subref{fig:flux_map}, \subref{fig:flux_map_1TeV}, and \subref{fig:flux_map_5TeV}.
Additionally, we show in panel~\subref{fig:flux_map_hires} a flux map smoothed with a Gaussian kernel of $0.07^\circ$ width, which corresponds approximately to the size of the point-spread function for the analysis configuration employed here.

Very strong $\gamma$-ray emission is observed from the known, nearby sources HESS~J1640$-$465 and HESS~J1641$-$463.
Turning to HESS~J1646$-$458, we observe that its spatial morphology is very complex.
Notably, the emission is not peaked at the position of Westerlund~1, but rather exhibits a structure resembling that of a shell, surrounding the stellar cluster.
This global structure is present in all of the displayed maps, and thus does not seem to vary with $\gamma$-ray energy.
On top of the large-scale structure, we identify peaks of the emission in the circular regions labelled `A' and `B', which correspond to regions with enhanced emission already found in \citet{HESS_Westerlund1_2012}, confirming these findings.
Additionally, we find a peak -- visible in particular in the flux map above~\SI{4.9}{\TeV} -- in region `C', which encompasses the two energetic pulsars PSR~J1648$-$4611 and PSR~J1650$-$4601, as well as an emission region extending beyond the shell-like structure, to the east of region~C.
For future reference, we provide the following source names for these regions: HESS~J1645$-$455 (region~A), HESS~J1647$-$465 (region~B), HESS~J1649$-$460 (region~C), and HESS~J1652$-$462 (emission east of region~C).
We stress, however, that we have found -- as will be detailed in the course of this paper -- no indications that the $\gamma$-ray emission in these regions is of a different origin than that of the rest of the emission, and that the regions should therefore not be interpreted as distinct sources.
Rather, the regions have been labelled in order to ease the discussion of the source morphology.

Because HESS~J1646$-$458 is located along the Galactic plane, and towards the inner Galaxy, it is safe to assume that diffuse $\gamma$-ray emission contributes to the observed signal to some degree.
This diffuse emission is produced by CRs that propagate freely within the Galactic disc, and can be due to bremsstrahlung or IC emission of CR electrons, or interactions of hadronic CRs with gas.
Due to its diffuse nature, the diffuse $\gamma$-ray emission from the Galaxy is challenging to measure directly, and while it has been detected over large scales in the TeV energy range \citep[e.g.,][]{HESS_Diffuse_2014,Tibet2021}, these measurements do not provide a good constraint for the level of diffuse emission in the region of HESS~J1646$-$458.
Therefore, in order to assess the possible contamination with diffuse emission of the $\gamma$-ray signal of HESS~J1646$-$458, we have used a prediction of the diffuse $\gamma$-ray flux based on the \textsc{Picard} CR propagation code \citep{Kissmann2014,Kissmann2015,Kissmann2017}.
This analysis is described in more detail in Appendix~\ref{sec:appendix_picard}, where we show in Fig.~\ref{fig:flux_maps_picard_subtracted} the same flux maps as in Fig.~\ref{fig:flux_maps}, but with the predicted flux due to diffuse emission subtracted.
We conclude that, while the Galactic diffuse emission likely contributes at a considerable level -- $\sim$24\% ($\sim$17\%/$\sim$8\%) above a threshold energy of \SI{0.37}{\TeV} (\SI{1}{\TeV}/\SI{4.9}{\TeV}), according to the \textsc{Picard} template --, it cannot explain the bulk of the $\gamma$-ray emission, and does not alter the source morphology in a significant way.
For these reasons, and because of the rather large uncertainties associated with any estimate of the Galactic diffuse emission in a particular region of the sky, we have performed the subsequent analysis without explicitly taking it into account, noting that none of the conclusions drawn in this paper are affected by this.

\begin{figure}
  \centering
  \includegraphics{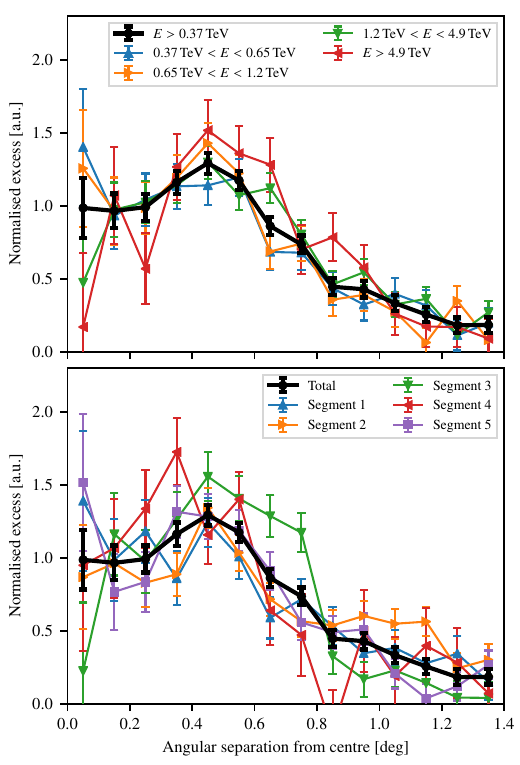}
  \caption{
    Radial excess profiles.
    The upper panel shows exposure-corrected excess counts per unit sky area for different energy bands, the lower panel for different segments as defined in Fig.~\ref{fig:sign_map_boxes} -- the black curve, showing the total excess above threshold in all segments, is the same in both panels.
    All profiles are normalised to equal area, to allow an easy comparison.
    The profiles have been computed with respect to a centre point at R.A.~$16^\mathrm{h}46^\mathrm{m}36^\mathrm{s}$, Dec.~$-46^\circ 01'12''$, slightly shifted from the Westerlund~1 position.
    In the calculation of the profiles, we discarded pixels within $0.6^\circ$ of the position of HESS~J1640$-$465.
  }
  \label{fig:radial_profiles}
\end{figure}

In order to further characterise the morphology of the emission -- and its apparent invariance with respect to energy -- we derived radial profiles of the observed excess.
Noting that Westerlund~1 is not located precisely at the centre of the shell-like structure, the profiles were computed not with respect to the stellar cluster position, but to a slightly shifted coordinate (R.A.~$16^\mathrm{h}46^\mathrm{m}36^\mathrm{s}$, Dec.~$-46^\circ 01'12''$), which corresponds to the barycentre of the $\gamma$-ray excess.
The asymmetry of the observed emission with respect to the cluster position could for example be caused by inhomogeneities in the ISM surrounding the stellar cluster.
Fig.~\ref{fig:radial_profiles} shows the profiles for different energy bands (upper panel), and for the five segments defined in Fig.~\ref{fig:sign_map_boxes} (lower panel).
The excess profiles:
\begin{enumerate}[(i)]
  \item confirm the shell-like structure, with a peak at around $0.5^\circ$ (corresponding to a projected distance of $\sim$\SI{34}{\pc} for a cluster distance of \SI{3.9}{\kpc}), followed by a relatively slow fall-off;
  \item exhibit the same shape in all energy bands, that is, they show no indications for an energy-dependent morphology of the excess;
  \item are also largely compatible between the five segments, with only minor small-scale deviations discernible.
\end{enumerate}
In order to reinforce the second point above, we carried out $\chi^2$ tests in which we compared the profile for each energy band with one computed using all events outside this band (thus ensuring statistically independent sets).
The results, listed in Table~\ref{tab:ebands}, show that each of the profiles is compatible with the total profile in terms of its shape within the statistical uncertainties.

\begin{table}
  \centering
  \caption{Morphological fit results.}
  \label{tab:ebands}
  \begin{tabular}{cccc}
      \hline\hline
      Energy range & Excess & $\chi^2/N_\mathrm{dof}$ & $p_{\chi^2}$\\
      (TeV) &  &  & \\\hline
      $>0.37$ & \num{15310\pm 440} & -- & --\\
      0.37 -- 0.65 & \num{5080\pm 300} & 12.2 / 14 & 59.0\%\\
      0.65 -- 1.2 & \num{3910\pm 230} & 16.2 / 14 & 30.3\%\\
      1.2 -- 4.9 & \num{5190\pm 220} & 20.9 / 14 & 10.3\%\\
      $>4.9$ & \num{1130\pm 80} & 19.5 / 14 & 14.6\%\\
      \hline
  \end{tabular}
  \tablefoot{
    `Excess' denotes the number of observed excess events within the white, dashed circle in Fig.~\ref{fig:sign_map_boxes}, excluding a circular region with $0.6^\circ$ radius around HESS~J1640$-$465.
    The $\chi^2$ values and corresponding $p$-values are a measure of the compatibility of the shape of the radial excess profile of this energy band with the total radial excess profile (see text for details).
  }
\end{table}

\subsubsection{Energy spectra}
\label{sec:spectra}

\begin{figure}
  \centering
  \includegraphics{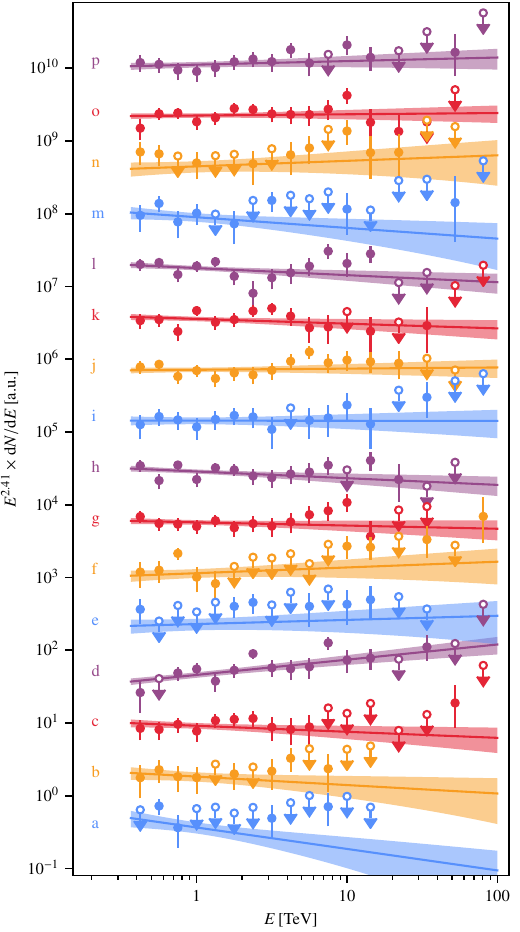}
  \caption{
    Comparison of signal region spectra.
    All spectra were divided (at a reference energy of \SI{1}{\TeV}) by a reference power-law spectrum with spectral index $\Gamma=2.41$, corresponding to the weighted average over all signal regions.
    Upper limits are at 95\% confidence level, and only two upper limits after the last significant (i.e., $>2\sigma$) flux point are shown.
  }
  \label{fig:box_spectra_comparison}
\end{figure}

\begin{figure}
  \centering
  \includegraphics{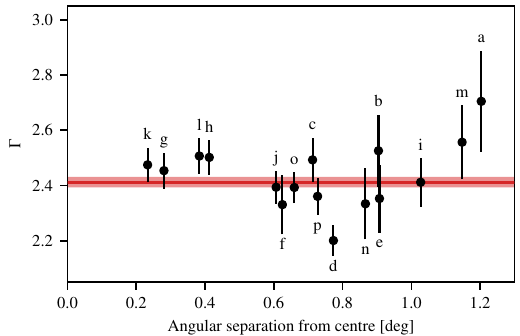}
  \caption{
    Spectral index $\Gamma$ for the signal regions a--p, as a function of angular separation between the centre point of each region and the centre point of the total emission (white circle in Fig.~\ref{fig:flux_maps}).
    The red line and band denote the weighted mean and uncertainty across all regions, $\langle \Gamma\rangle = 2.41\pm 0.02$, respectively.
  }
  \label{fig:box_index_vs_distance}
\end{figure}

The complex morphology of HESS~J1646$-$458 prohibits a simple extraction of an energy spectrum by means of modelling the emission with a single spatial model.
We therefore considered the 16~signal regions indicated in Fig.~\ref{fig:sign_map_boxes} and extracted energy spectra for each of these regions.
To obtain the spectra, we fitted \citep[using a forward-folding likelihood fit;][]{Piron2001} a power-law (PL) model,
\begin{equation}\label{eq:pl}
  \frac{\mathrm{d}N}{\mathrm{d}E} = \phi_0\cdot \left(\frac{E}{E_0}\right)^{-\Gamma}\quad,
\end{equation}
with $E_0=\SI{1}{\TeV}$ kept fixed in the fit, to the observed $\gamma$-ray excess in each signal region and subsequently derived flux points based on this fitted model.
Table~\ref{tab:box_stats} lists the observed $\gamma$-ray excess as well as the fitted power-law model parameters.
A comparison of the shapes of the energy spectra is provided in Fig.~\ref{fig:box_spectra_comparison}.
Finally, Fig.~\ref{fig:box_index_vs_distance} shows the fitted spectral index for each signal region as a function of its angular separation from Westerlund~1.

\begin{table*}
  \centering
  \caption{Spectral analysis results for signal regions.}
  \label{tab:box_stats}
  \begin{tabular}{cS[table-align-text-post=false]S[table-align-text-post=false]S[table-align-text-post=false]ccS[table-align-text-post=false]}
    \hline\hline
    Signal region & {Excess events} & {Significance} & {Significance} & $\phi_0$ & $\Gamma$ & $\sqrt{\Delta\mathrm{TS}}$\\
     & & & {$(E>\SI{4.9}{\TeV})$} & $(10^{-13}\,\mathrm{TeV}^{-1}\mathrm{cm}^{-2}\mathrm{s}^{-1})$ & & \\\hline
    a & 396.1 & 5.3$\sigma$ & 0.9$\sigma$ & $3.76\pm 0.66$ & $2.71\pm 0.18$ & 5.9\\
    b & 454.9 & 5.6$\sigma$ & 1.7$\sigma$ & $4.34\pm 0.64$ & $2.53\pm 0.13$ & 7.5\\
    c & 901.8 & 10.3$\sigma$ & 2.8$\sigma$ & $6.33\pm 0.58$ & $2.49\pm 0.08$ & 12.3\\
    d & 1014.0 & 10.8$\sigma$ & 7.7$\sigma$ & $6.66\pm 0.58$ & $2.20\pm 0.06$ & 16.1\\
    e & 430.7 & 4.7$\sigma$ & 2.9$\sigma$ & $2.84\pm 0.51$ & $2.35\pm 0.12$ & 6.7\\
    f & 648.9 & 7.7$\sigma$ & 4.0$\sigma$ & $4.60\pm 0.64$ & $2.33\pm 0.11$ & 10.0\\
    g & 1238.5 & 13.5$\sigma$ & 6.0$\sigma$ & $7.41\pm 0.54$ & $2.45\pm 0.07$ & 16.1\\
    h & 1409.2 & 14.5$\sigma$ & 4.6$\sigma$ & $8.14\pm 0.54$ & $2.50\pm 0.06$ & 17.3\\
    i & 653.4 & 9.0$\sigma$ & 4.0$\sigma$ & $6.65\pm 0.71$ & $2.41\pm 0.09$ & 11.4\\
    j & 1229.0 & 14.0$\sigma$ & 6.8$\sigma$ & $9.07\pm 0.63$ & $2.39\pm 0.06$ & 17.7\\
    k & 1246.4 & 13.2$\sigma$ & 3.6$\sigma$ & $7.73\pm 0.54$ & $2.48\pm 0.06$ & 16.5\\
    l & 1405.7 & 14.1$\sigma$ & 6.3$\sigma$ & $7.95\pm 0.54$ & $2.51\pm 0.06$ & 16.9\\
    m & 469.5 & 6.8$\sigma$ & 1.7$\sigma$ & $5.40\pm 0.73$ & $2.56\pm 0.13$ & 8.2\\
    n & 415.4 & 5.1$\sigma$ & 3.5$\sigma$ & $3.49\pm 0.62$ & $2.33\pm 0.13$ & 7.4\\
    o & 1259.2 & 14.1$\sigma$ & 5.9$\sigma$ & $8.23\pm 0.57$ & $2.39\pm 0.06$ & 17.7\\
    p & 996.7 & 10.5$\sigma$ & 4.0$\sigma$ & $6.29\pm 0.55$ & $2.36\pm 0.07$ & 14.7\\
    \hline
  \end{tabular}
  \tablefoot{
    See Fig.~\ref{fig:sign_map_boxes} for the definition of the signal regions.
    Significance values were computed following \citet{Li1983}, assuming a perfect knowledge of the background.
    $\phi_0$ and $\Gamma$ are the best-fit parameter values of the power-law fit for each region (cf.~Eq.~(\ref{eq:pl})).
    $\sqrt{\Delta\mathrm{TS}}$ denotes the square root of the difference in test statistic ($\mathrm{TS}=-2\ln(\mathcal{L})$) between the best-fit power-law model and the null hypothesis (corresponding to $\phi_0=0$).
  }
\end{table*}

The spectra in the signal regions are remarkably similar to each other, both in terms of the fitted power-law indices (which show no dependence on the separation of the region from the centre, see Fig.~\ref{fig:box_index_vs_distance}) and the shape of the spectra as indicated by the extracted flux points (see Fig.~\ref{fig:box_spectra_comparison}).
The only significant deviation is observed in region `d', where the fitted power-law index deviates from the average of all other regions by $\sim 4\sigma$.
We have not been able to identify an issue in the analysis procedure that could explain this deviation, and conclude that it either indicates that the spectrum of the emission in this region is indeed harder, or that it is an unexpectedly large statistical fluctuation.
The similarity of the spectra supports our previous finding of a lack of energy-dependent morphology of HESS~J1646$-$458, and motivates the extraction of a combined energy spectrum.

\begin{figure}
  \centering
  \includegraphics{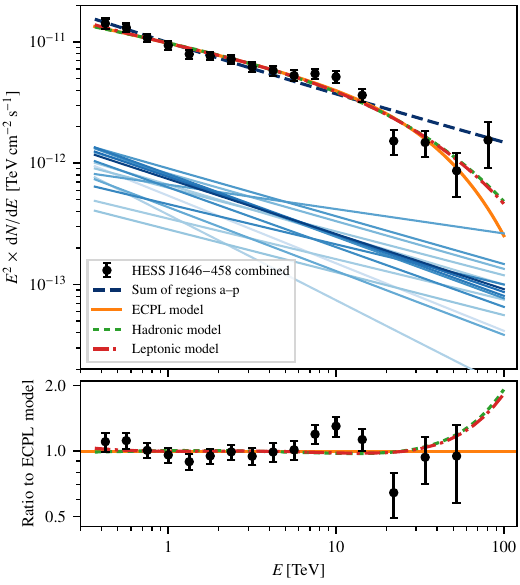}
  \caption{
    Combined energy spectrum.
    The black data points correspond to the entire emission of HESS~J1646$-$458; the solid orange, dashed green, and dashed-dotted red lines show the result of fitting a power law with exponential cut-off (ECPL), a hadronic ($pp$) model, and a leptonic (IC) model, respectively, to these points.
    Fitted power-law models for each region a--p are displayed by solid blue lines (with darker shades indicating closer proximity to Westerlund~1), while the dashed blue line denotes their sum.
    All power-law spectra are plotted up to \SI{100}{\TeV} for better visibility, however, the observed $\gamma$-ray excess is not significant up to this energy in any of the sub-regions.
    The bottom panel shows the ratio to the ECPL model; note that the last flux point (with a ratio to the ECPL model of $\sim$3.7$\pm$1.5) is beyond the vertical axis scale.
  }
  \label{fig:combined_spectrum}
\end{figure}

We computed combined flux points for HESS~J1646$-$458 by adding up the flux points from all 16~square regions, where we have used the best-fit flux for each point and region also in cases where an upper limit was previously derived.
The result is shown in Fig.~\ref{fig:combined_spectrum}.
Besides statistical uncertainties, the displayed error bars contain a systematic uncertainty arising from the applied background model.
The systematic uncertainty is of the same magnitude as the statistical one at the lowest energies considered here, and quickly becomes negligible at higher energies.
The combined flux points clearly show that the $\gamma$-ray emission of HESS~J1646$-$458 extends to at least several tens of TeV.
A comparison with the spectra of the individual signal regions (blue lines in Fig.~\ref{fig:combined_spectrum}) shows that the total spectrum is not dominated by any single signal region across the entire energy range.

In order to characterise the combined spectrum, we fitted several spectral models to the derived flux points.
A simple PL model (cf.\ Eq.~\ref{eq:pl}) does not lead to a satisfactory fit ($p=0.06\%$).
The solid orange line in Fig.~\ref{fig:combined_spectrum} shows the result for a power law with exponential cut-off (ECPL),
\begin{equation}\label{eq:ecpl}
  \frac{\mathrm{d}N}{\mathrm{d}E} = \phi_0\cdot \left(\frac{E}{E_0}\right)^{-\Gamma}\cdot \exp\left(-\frac{E}{E_c}\right)\quad,
\end{equation}
for which we obtained (with $E_0=\SI{1}{\TeV}$ kept fixed in the fit) $\phi_0=\SI{1.00(03)E-11}{\per\TeV\per\square\cm\per\second}$, $\Gamma=2.30\pm 0.04$, and $E_c=(44^{+17}_{-11})\,\mathrm{TeV}$.
This corresponds to a total $\gamma$-ray luminosity of HESS~J1646$-$458 between the threshold energy of~\SI{0.37}{\TeV} and \SI{100}{\TeV} of $L_\gamma\sim\num{9e34}\,(d/\SI{3.9}{\kpc})^2\,\si{\erg\per\second}$, where $d$ is the assumed distance to the source.
We note that, while the ECPL model yields an acceptable fit ($p=6.3\%$), the high-energy flux points do not provide a clear indication of an exponential cut-off to the spectrum.
Thus, while the energy spectrum of HESS~J1646$-$458 clearly extends to several tens of TeV, its maximum energy cannot be determined reliably with the analysis presented here, and may conceivably lie beyond \SI{100}{TeV}.
However, the last flux point should not be regarded as a clear indication of an upward trend in the spectrum, as it may be afflicted by unknown systematic uncertainties in the high-energy response of the system, which is difficult to calibrate.

Assuming the $\gamma$-ray emission to be generated in collisions of CR protons with ambient matter, we also fitted a primary proton spectral model (of the same form as defined in Eq.~\ref{eq:ecpl}) to the $\gamma$-ray flux points, employing the \textsc{Naima} software package \citep{Zabalza2015} for this task.
For the parameters of the primary proton spectrum we obtained a normalisation (at $E_0=\SI{1}{\TeV}$) of $\phi_0^p=\num{1.28(17)E+38}\,(d/\SI{3.9}{\kpc})^2\,(n/\SI{1}{\per\cubic\cm})^{-1}\,\si{\per\eV}$ (with $n$ the assumed density of the ambient matter), a spectral index of $\Gamma_p=2.33\pm 0.06$, and a cut-off energy of $E_c^p=(400^{+250}_{-130})\,\mathrm{TeV}$; the dashed green line in Fig.~\ref{fig:combined_spectrum} displays the corresponding $\gamma$-ray spectrum.
The 95\% confidence level lower limit on the proton spectrum cut-off energy is $E_c^p > \SI{214}{\TeV}$.
Extrapolating the proton spectrum down to an energy of \SI{1}{\GeV}, the required energy in primary protons is $W_p\sim \num{6e51}\,(d/\SI{3.9}{\kpc})^2\,(n/\SI{1}{\per\cubic\cm})^{-1}\,\si{\erg}$.

In a similar manner, now adopting with \textsc{Naima} a leptonic framework in which the $\gamma$-ray emission is produced through IC scattering of CR electrons, we also fitted a primary electron spectrum to the HESS~J1646$-$458 flux points (again assuming an ECPL model).
Besides the cosmic microwave background, we used as target radiation fields an infrared field ($T_\mathrm{IR}=\SI{26}{K}$, $\rho_\mathrm{IR}=\SI{0.74}{\eV\per\cubic\cm}$) and an optical field ($T_\mathrm{opt}=\SI{2400}{K}$, $\rho_\mathrm{opt}=\SI{1.4}{\eV\per\cubic\cm}$) as predicted by the \citet{Popescu2017} model, as well as an additional field representing diffuse star light from the stellar cluster.
For the latter, we assumed $T_\mathrm{SC}=\SI{40000}{K}$ \citep{Crowther2006} and derived\footnote{
  To derive the energy density, we used $\rho_\mathrm{SC}=L\,/\,(4\pi r^2 c)$, where $L$ is the total cluster luminosity and $r$ the distance from the cluster.
  For a wind efficiency $\eta=\dot{M}v_\mathrm{w}\,/\,(L/c)\simeq 1$ \citep{Vink2012}, $L=2\,(v_\mathrm{w}/c)^{-1}L_\mathrm{w} \sim \SI{2e41}{\erg\per\second}$ for the parameter values assumed in this work.
  Adding up the luminosities of OB supergiant and Yellow Hypergiant stars listed in \citet{Clark2005} yields $L\sim \SI{6e40}{\erg\per\second}$, which is consistent with our estimate.
  A slight reduction of the IC emission due to the cluster radiation field is expected because of its anisotropy, which we have not taken into account.
  However, since the contribution of this component is suppressed due to the Klein-Nishina effect, the modification of the total IC emission is negligible.
}
an energy density of $\rho_\mathrm{SC}\sim \SI{30}{\eV\per\cubic\cm}$ at a distance of \SI{34}{\pc} from the cluster, which approximately corresponds to the distance at which the radial $\gamma$-ray excess profile peaks (cf.\ Fig.~\ref{fig:radial_profiles}).
In this case, the best-fit parameters are $\phi_0^e=\num{4.7(5)E+35}\,(d/\SI{3.9}{\kpc})^2\,\si{\per\eV}$, $\Gamma_e=2.97\pm 0.07$, and $E_c^e=(180^{+200}_{-70})\,\mathrm{TeV}$, with a 95\% confidence level lower limit on $E_c^e$ of \SI{87}{\TeV} -- the resulting $\gamma$-ray spectrum is shown by the red, dashed-dotted line in Fig.~\ref{fig:combined_spectrum}.
Electrons down to energies of $\sim$\SI{0.4}{\TeV} contribute to the $\gamma$-ray emission detected with H.E.S.S..
Assuming that the spectrum of primary electrons extends down to \SI{0.1}{\TeV}, we obtained a total required energy in primary electrons of $W_e\sim \num{7.2e48}\,(d/\SI{3.9}{\kpc})^2\,\si{\erg}$.
Dividing the required energy by the (energy-dependent) cooling time due to IC scattering \citep{Hinton2009} off the different target fields then yields an estimate for the minimum total required power of $L_e > \num{4.1e35}\,(d/\SI{3.9}{\kpc})^2\,\si{\erg\per\second}$.
The required power is larger if cooling due to synchrotron emission plays a sizeable role, for example, in a magnetic field with $B=\SI{10}{\micro\gauss}$, $L_e > \num{1.7e36}\,(d/\SI{3.9}{\kpc})^2\,\si{\erg\per\second}$.

The chosen analysis method and tools in principle also enable a three-dimensional modelling of the $\gamma$-ray emission detected with H.E.S.S., that is, to decompose the emission into several components with separate energy spectra and morphological models.
However, owing to the complex structure of the emission, a rather complicated model with multiple distinct components is required to obtain a satisfactory description, and its interpretation depends strongly on the chosen models for each of the components.
Considering furthermore the similarity of energy spectra extracted in the signal regions, we do not present such a modelling here.

\subsection{Analysis of radio line data}
\label{sec:radio}

Attributing the $\gamma$-ray emission to interactions of accelerated cosmic-ray nuclei requires sufficiently dense target material.
The presence of such target material can be estimated using radio observations, in particular of the \SI{21}{\cm} H~I emission line -- indicating neutral, atomic hydrogen -- and of the CO~($J$=1--0) transition, which is a tracer for dense clouds of molecular hydrogen, H$_2$ \citep{Heyer2015}.
We have therefore used the H~I Southern Galactic Plane Survey \citep[SGPS;][]{McClureGriffiths2005} and the Mopra Southern Galactic Plane CO Survey \citep{Braiding2018} to investigate the amount of hydrogen gas in the vicinity of Westerlund~1.
The analysis was repeated using the CO data from the (lower-resolution) survey by \citet{Dame2001} instead of the Mopra CO data, leading to consistent results.

\begin{figure}
  \centering
  \includegraphics{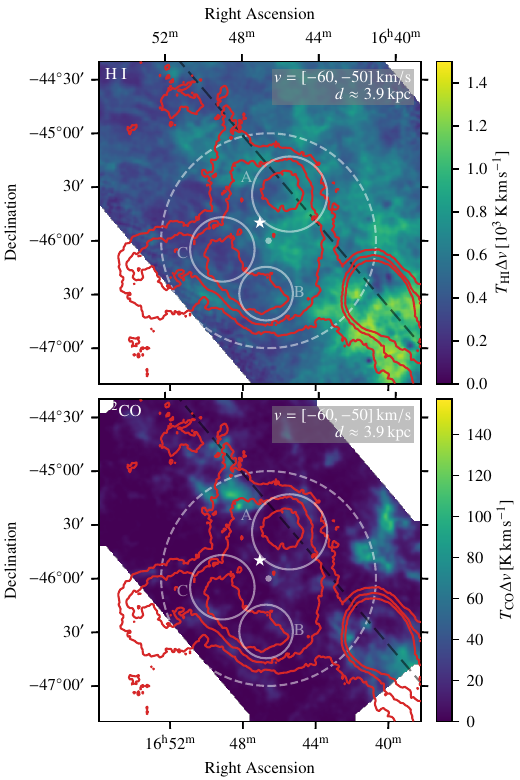}
  \caption{
    Maps showing H I emission \citep{McClureGriffiths2005} (\emph{top panel}) and $^{12}$CO emission \citep{Braiding2018} (\emph{bottom panel}) in the Westerlund~1 region.
    Both maps display the emission for an interval in velocity with respect to the local standard of rest of $v=[-60,-50]\,\si{\km\per\second}$, which approximately corresponds to a distance of \SI{3.9}{\kpc}.
    The position of Westerlund~1 is marked by the white star symbol and the grey, dashed line shows the Galactic plane.
    The transparent, white circle marker denotes the centre point with respect to which the radial CR density profiles in Fig.~\ref{fig:cr_dens_profile_shifted} have been computed; the dashed white line displays a circle with radius $1^\circ$ -- up to which the profiles have been computed -- around this point. 
    The red lines are contour lines of the flux map shown in Fig.~\ref{fig:flux_map}.
    Regions A, B, and C are the same as in Fig.~\ref{fig:flux_maps}.
  }
  \label{fig:hi_co_map_dist_3d9}
\end{figure}

The radio data analysis is hampered by the uncertainty on the distance to Westerlund~1.
We show in Fig.~\ref{fig:hi_co_map_dist_3d9} the H~I and CO maps for an interval in velocity with respect to the local standard of rest of $v=[-60,-50]\,\si{\km\per\second}$, which corresponds to the distance of \SI{3.9}{\kpc} that we adopted for this paper \citep{Kothes2007}.
As some measurements indicate smaller distances, maps for two correspondingly chosen alternative velocity intervals, $v=[-48.5, -38.5]\,\si{\km\per\second}$ ($d\approx \SI{3.3}{\kpc}$) and $v=[-37, -27]\,\si{\km\per\second}$ ($d\approx \SI{2.7}{\kpc}$), are shown in Appendix~\ref{sec:appendix_radio}.

We find that the gas indicated by the radio observations at a distance of $\sim$\SI{3.9}{\kpc} shows no spatial correlation with the $\gamma$-ray emission that we observe with H.E.S.S..
In fact, both the H~I and CO maps indicate a particularly low atomic and molecular gas density in the circular regions~B and~C, which are bright in $\gamma$ rays.
Using an H~I intensity-mass conversion factor of $X_\mathrm{HI}=\num{1.823e18}\,\mathrm{cm}^{-2}\,/\,(\si{\kelvin\km\per\second})$ \citep{Rohlfs2004}, we obtain for a circular region with radius $1.1^\circ$, centred on Westerlund~1, a total enclosed mass as indicated from H~I of $M_\mathrm{H~I,Wd1}=\num{1.3e5}\,M_\odot$.
This translates into an average density of $n_\mathrm{H~I,Wd1}=\SI{3.2}{\per\cubic\cm}$.\footnote{
  \citet{HESS_Westerlund1_2012} derived, for a similar region, a much smaller value of $n_\mathrm{H~I}=\SI{0.24}{\per\cubic\cm}$.
  We attribute this to the usage of an erroneous formula in that paper.
}
Similarly, from the CO data we get\footnote{
  Due to the more indirect nature of the estimate, the CO-to-H$_2$ conversion factor, $X_\mathrm{CO}$, is less well constrained than $X_\mathrm{H~I}$.
  Here we used $X_\mathrm{CO}=\num{1.5e20}\,\mathrm{cm}^{-2}\,/\,(\si{\kelvin\km\per\second})$, which \citet{FermiLAT2012} indicate as an appropriate value for the galactocentric radius of Westerlund~1, $R\approx \SI{4.7}{\kpc}$ for a distance of \SI{3.9}{\kpc} (see their Fig.~25).
  This value is also within the range of $(1.4-2.6)\times 10^{20}\,\mathrm{cm}^{-2}\,/\,(\si{\kelvin\km\per\second})$ recommended by \citet{Bolatto2013}.
  We have neglected the possible contribution of $^4$He, which could increase the mass estimate by $\sim$25\%.
}
$M_\mathrm{CO,Wd1}=\num{4.3e5}\,M_\odot$ and $n_\mathrm{CO,Wd1}=\SI{10.5}{\per\cubic\cm}$, where $n_\mathrm{CO}$ is the equivalent density for atomic hydrogen and can thus be directly compared to $n_\mathrm{H~I}$.
We stress, however, that in particular the molecular material indicated by the CO observations is not distributed homogeneously inside this region, but rather concentrated in smaller-scale clouds.
For instance, in the CO cloud located in the Northern part of the H.E.S.S.\ emission region (cf.\ Fig.~\ref{fig:hi_co_map_dist_3d9}), we find -- assuming a spherical distribution of the gas -- a density of $n_\mathrm{CO,cloud}\sim \SI{190}{\per\cubic\cm}$.

\begin{figure}
  \centering
  \subfigure[]{
    \includegraphics{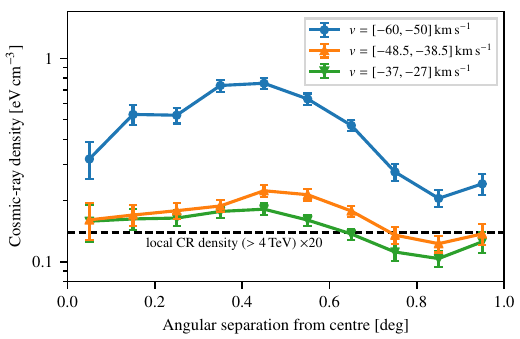}
    \label{fig:cr_dens_profile_shifted}
  }
  \subfigure[]{
    \includegraphics{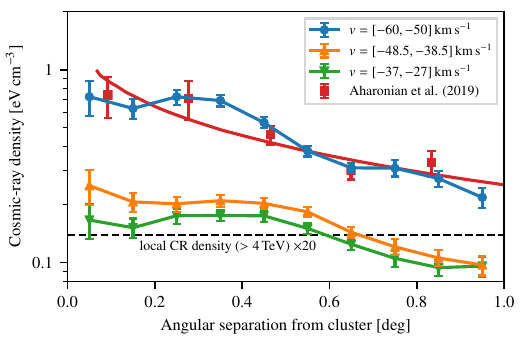}
    \label{fig:cr_dens_profile_cluster}
  }
  \caption{
    Cosmic-ray density profiles above \SI{4}{\TeV} for different velocity intervals.
    The profiles in the upper plot have been computed with respect to the centre point shown by the circle marker in Fig.~\ref{fig:hi_co_map_dist_3d9} (the same as for the excess profiles displayed in Fig.~\ref{fig:radial_profiles}), whereas the profiles in the lower plot have been computed with respect to the position of Westerlund~1.
    The error bars denote the (statistical) uncertainty of the measured $\gamma$-ray flux only, and in particular do not reflect systematic uncertainties related to the gas distribution.
    The CR density at Earth, indicated for comparison, was computed from the all-particle CR spectrum as modelled in \citet{Breuhaus2022} (model `A').
    The data points taken from \citet{Aharonian2019}, shown in the lower plot, have been derived using a velocity interval of $v=[-60,-50]\,\si{\km\per\second}$, and should thus be compared to the blue curve with circle markers.
    The red, solid line, from the same paper, shows a profile corresponding to a $1/r$-distribution of the CRs, where $r$ is the distance to the cluster.
    Both show the density above a slightly higher threshold energy of \SI{10}{\TeV}.
  }
  \label{fig:cr_dens_profiles}
\end{figure}

Following \citet{Aharonian2019}, we also derived a radial profile of the CR density, assuming that the $\gamma$-ray emission is produced in interactions of hadronic CRs with the gas indicated by the H~I and CO data.
Having used the measured $\gamma$-ray flux above our threshold energy of \SI{0.37}{\TeV} to compute the profiles, they indicate the density of CRs above an energy about ten times higher, that is, above $\sim$\SI{4}{\TeV}.
Density profiles for all three considered velocity intervals are shown in Fig.~\ref{fig:cr_dens_profiles}, where we have used for the profiles in panel \subref{fig:cr_dens_profile_shifted} the same shifted centre point as for the radial excess profiles (cf.\ Fig.~\ref{fig:radial_profiles}), and for the profiles in panel \subref{fig:cr_dens_profile_cluster} -- for comparison with \citet{Aharonian2019}~-- the position of Westerlund~1 as centre point.
Regarding first Fig.~\ref{fig:cr_dens_profile_shifted}, for the interval corresponding to the nominal distance of $d\approx \SI{3.9}{\kpc}$ (blue curve with circle markers), we observe a distinct peak at a radial distance from the centre of $\sim$$0.4^\circ$, corresponding to the peak observed in the excess profiles (cf.\ Fig.~\ref{fig:radial_profiles}) at about the same distance.
When choosing the cluster position as centre point, the peak is smeared out, leading to a plateau at small radii.
The profile is then compatible with that derived in \citet{Aharonian2019}, and showing a gradual decline of the density moving away from the cluster.
We stress, however, that the $\gamma$-ray emission is not radially symmetric around Westerlund~1, rendering the interpretation of radial profiles computed with respect to this position difficult.
Our profiles for the alternative velocity intervals (i.e., distances; orange and green curves with triangle markers in Fig.~\ref{fig:cr_dens_profiles}) exhibit approximately the same shape as the profile for the nominal distance, but with a less distinct peak, and lower overall density.
This corresponds to the higher density of hydrogen gas observed for these intervals (cf.\ Figs.~\ref{fig:hi_co_map_dist_3d3},~\ref{fig:hi_co_map_dist_2d7}).
Finally, we note that the shapes of the obtained radial density profiles should be interpreted with care, owing to systematic uncertainties associated with the determination of the gas distributions.
For instance, it is conceivable that a considerable amount of the CO molecules have been photodissociated by the intense ultraviolet radiation from Westerlund~1, implying that the CO line emission would no longer be an accurate tracer of molecular hydrogen gas \citep[e.g.,][]{Wolfire2010}.

\section{Discussion}
\label{sec:discussion}

In this section, we will discuss in turn the various possible sites for the acceleration of CRs that could be responsible for the observed $\gamma$-ray emission.
Evidently, we can include in this discussion only those objects that have been identified through observations at other wavelengths.
While we consider the association of the $\gamma$-ray emission with one of these objects most likely, so far undiscovered objects (e.g. pulsars or SNRs) could also contribute to it.

\subsection{Objects not connected to Westerlund~1}

\subsubsection{4U~1642--45}
The LMXB 4U~1642$-$45 lies close to the peak in $\gamma$-ray emission observed in region~A (cf.\ Fig.~\ref{fig:flux_maps}).
However, despite this positional coincidence, we consider an association of 4U~1642$-$45 with HESS~J1646$-$458 -- or even the emission observed just in region~A -- highly unlikely: LMXBs are not known emitters of $\gamma$ rays in the TeV energy range, in fact, \citet{Kantzas2022} have recently shown that even the upcoming Cherenkov Telescope Array (CTA) will be able to detect LMXB outbursts only under favourable circumstances.
Moreover, the observed emission is spatially extended, which would be unexpected if it is produced inside a transient jet.
Lastly, \citet{HESS_Westerlund1_2012} reported no temporal variations of the observed $\gamma$-ray flux, again disfavouring an association of HESS~J1646$-$458 with 4U~1642$-$45.
While the deviation in spectral index observed in signal region~`d' (cf.\ Sect.~\ref{sec:spectra}), which partly coincides with the circular region~A, is intriguing, the lack of a plausible association leads us to conclude that it is likely a statistical fluctuation, or due to an unidentified, hard-spectrum source contributing to the emission in region~d (but not to the entire region~A, as the signal regions~c, g, and h, which also overlap with region~A, do not show deviating spectral indices).

\subsubsection{PSR~J1648--4611 \& PSR~J1650--4601}
The two energetic pulsars PSR~J1648$-$4611 (with spin-down power $\dot{E}=\SI{2.1e35}{\erg\per\s}$) and PSR~J1650$-$4601 ($\dot{E}=\SI{2.9e35}{\erg\per\s}$) \citep{Manchester2005} are located within region~C (cf.\ Fig.~\ref{fig:flux_maps}), where we observe enhanced $\gamma$-ray emission.
Because pulsar wind nebulae (PWNe) represent a large fraction of known Galactic $\gamma$-ray sources \citep[e.g.,][]{HESS_HGPS_2018}, it seems likely that high-energy electrons provided by one (or both) of the two pulsars contribute to the $\gamma$-ray emission observed in region~C.
This remains true even though neither of the pulsar locations fully coincides with the peak of the emission, as it is not uncommon that $\gamma$-ray PWNe are observed offset from their respective pulsars.
The detection of diffuse, hard X-ray emission from the vicinity of PSR~J1648$-$4611 by \citet{Sakai2013} adds further support for this scenario.
However, while the 4FGL-DR2 catalogue \citep{FermiLAT2020,FermiLAT2020a} contains sources associated with PSR~J1648$-$4611 and PSR~J1650$-$4601, their $\gamma$-ray emission is detected as pulsed and exhibits a very steep spectrum above \SI{10}{\GeV}, implying that these sources are directly connected to the pulsars rather than their putative PWNe.

On the other hand, viewing the entire emission of HESS~J1646$-$458 as resulting from one of the two pulsars would imply an unusually large PWN.
For instance, at the distance of PSR~J1648$-$4611 of \SI{5.7}{\kpc} \citep{Kramer2003}, the emission region spans $\sim$\SI{200}{\pc}, which is twice the size of the largest known $\gamma$-ray PWN, HESS~J1825$-$137 \citep{HESS_1825_2019}.
For such an extended nebula, one would expect a considerable loss of energy due to synchrotron cooling of the electrons when they propagate towards the edges of the nebula, leading to softer $\gamma$-ray spectra in these regions (or, equivalently, energy-dependent morphology).
Because we observe very similar energy spectra across the entire source, and no energy-dependent morphology, we conclude that a PWN powered by PSR~J1648$-$4611 or PSR~J1650$-$4601 cannot explain the entire $\gamma$-ray emission.

\subsection{Distinct acceleration sites within Westerlund~1}
Having established that other known objects in the region cannot explain the bulk of $\gamma$-ray emission from HESS~J1646$-$458, we assume next that the CRs producing the emission are accelerated at one or multiple sites within the stellar cluster Westerlund~1.
Various scenarios can be considered, and will be discussed in this section.

\subsubsection{CXOU J164710.2--455216}
At first sight, the magnetar CXOU~J164710.2$-$455216 -- the only known stellar remnant inside the cluster -- may be suspected to power a $\gamma$-ray PWN that could potentially be associated to HESS~J1646$-$458.
However, its measured period and period derivative \citep[$P=\SI{10.6}{\second},\,\dot{P}=\SI{9.2e-13}{\second\per\second}$;][]{Israel2007} imply a rotational spin-down power of only $\dot{E}=\SI{3e31}{\erg\per\second}$, which is orders of magnitude lower than for any of the pulsars associated with PWN detected at TeV energies with H.E.S.S.\ \citep{HESS_PWNpop_2018}.
Even though the measured X-ray luminosity of $L_X\approx \SI{3e33}{\erg\per\second}$ \citep{Muno2006a} exceeds the rotational spin-down power, implying another source of energy (presumably connected to the high magnetic field of the magnetar), it still appears very unlikely that CXOU~J164710.2$-$455216 would be able to sustain the observed $\gamma$-ray emission.
Additionally, as is the case for PSR~J1648$-$4611 and PSR~J1650$-$4601, the spatial extent of HESS~J1646$-$458 and the lack of energy-dependent morphology disfavour an association of the $\gamma$-ray emission with CXOU~J164710.2$-$455216.

\subsubsection{Supernova remnants}
The existence of CXOU~J164710.2$-$455216 implies that at least one supernova (SN) explosion took place within Westerlund~1.
However, given the abundance of massive stars in Westerlund~1, and its age of several Myr, it seems certain that many more SNe have occurred already.
Indeed, \citet{Muno2006} have argued that the number of SNe during the last $\sim$\SI{1}{\mega\year} could be as high as 80--150, attributing the lack of identified SNRs to a cavity in the interstellar medium (ISM), excavated by the stellar cluster.
Without detailed knowledge about the progenitor mass of CXOU~J164710.2$-$455216, or the SN rate in Westerlund~1, it is not straightforward to estimate the energy output that SNRs in the stellar cluster may provide.
Nevertheless, assuming a canonical value for the kinetic energy released per SN of \SI{e51}{\erg} -- possibly more in the case of CXOU~J164710.2$-$455216, if its progenitor was indeed as heavy as $40\,M_\odot$ \citep{Muno2006a} -- it seems plausible that one or several SNRs in Westerlund~1 could be responsible for the $\gamma$-ray emission in terms of the required energetics.
For instance, in a hadronic scenario, assuming a density of the ambient matter of $5\,m_\mathrm{H}\,\si{\per\cubic\cm}$ gives a required energy in protons of \SI{1.2e51}{\erg} (cf.\ Sect.~\ref{sec:spectra}).
It would take about 10~SNRs to reach this energy if the conversion efficiency from kinetic energy into CRs is $\sim$10\%.

\subsubsection{SN-wind and wind-wind interactions}
\label{sec:wind_wind}
Another, related, possibility for the acceleration of CRs inside Westerlund~1 are interactions of SN shocks with winds of massive stars in the cluster, or interactions between the winds of several stars.
Indeed, for SN shocks interacting with fast stellar winds, the efficiency for CR acceleration can be as high as 30\% \citep{Bykov2020}, and CR acceleration up to $\geq\SI{40}{\PeV}$ has been conjectured \citep{Bykov2015}.
Colliding winds of massive stars are also known to produce non-thermal emission \citep{Eichler1993,Reimer2006}, and several searches for $\gamma$-ray emission from colliding wind binaries have been performed in the past \citep[e.g.][]{Werner2013,Pshirkov2016,MartiDevesa2021}.
A well-known example is provided by the colliding wind binary $\eta$ Car, which has indeed been detected up to $\sim$\SI{100}{\GeV} with the \emph{Fermi}-LAT \citep{AGILE2009,FermiLAT2010}, and even up to $\sim$\SI{400}{\GeV} with H.E.S.S.\ \citep{HESS_EtaCar_2020}.
These considerations strengthen the conclusion that $\gamma$-ray emission at the level observed with H.E.S.S.\ can in principle be produced by CRs accelerated at shock fronts within Westerlund~1.

However, for the scenario of a central CR source, it is important to also take into account the propagation of the CRs into the region where we observe $\gamma$-ray emission with H.E.S.S..
In this case, the non-observation of a peak in the $\gamma$-ray emission at the position of Westerlund~1 and the large extent of HESS~J1646$-$458 essentially rule out the leptonic scenario.
In a hadronic scenario, the complex morphology of HESS~J1646$-$458 may in principle be attributable to the distribution of target material -- although a clear correlation of the $\gamma$-ray emission with gas clouds as indicated by H~I and CO observations is lacking, and the inferred CR density does not peak at the centre for any of the considered distances to the source (cf.\ Fig.~\ref{fig:cr_dens_profile_shifted}), as would be expected for a steady CR injection there (see \citeauthor{Aharonian2019} \citeyear{Aharonian2019}, but also \citeauthor{Bhadra2022} \citeyear{Bhadra2022}).
Possible options to alleviate this problem could be the presence of ``dark'' gas that is not traced by H~I or CO \citep[e.g.,][]{Wolfire2010}, or the assumption that the CRs were provided by a recent impulsive event (e.g., the SN explosion of the CXOU~J164710.2$-$455216 progenitor star and/or other recent SNe), rather than being injected quasi-continuously over the lifetime of the cluster.
Another relevant constraint comes from the maximum energy of the observed $\gamma$ rays, which implies the presence of CRs with energies of several hundred~TeV throughout the emission region.
If the acceleration sites are located exclusively within the compact cluster, particles must pass through the wind zone.
Taking reasonable limits on the diffusion properties, adiabatic losses in the radial wind are unavoidable.
CRs that were injected at the cluster and have propagated to a distance $R$ within the wind region would therefore need to be produced with maximum energy $(R\,/\,R_\mathrm{Wd1})^{2/3}$ times larger \citep{Longair1992}, where $R_\mathrm{Wd1}\sim \SI{1}{\pc}$ is the radius of Westerlund~1.
For the nominal distance of \SI{3.9}{\kpc}, we observe a peak in the $\gamma$-ray emission at $R\sim \SI{34}{\pc}$, implying a need of multi-PeV CRs within Westerlund~1.
The H.E.S.S.\ observations thus provide a valuable constraint for theoretical models of particle acceleration processes within a compact stellar cluster \citep{Bykov2020}.

\subsection{Acceleration by Westerlund~1 as a whole}
Finally, there is the possibility that effects due to the entire stellar cluster provide the means for efficient CR acceleration.
In particular, we consider in the following two scenarios related to the collective cluster wind, in which the CR acceleration predominantly takes place outside of the actual stellar cluster itself.

\subsubsection{Turbulence in a superbubble}
Due to the powerful collective cluster wind, as well as many SN explosions, massive young stellar clusters are thought to create ``superbubbles'', that is, large cavities in the ISM, extending much beyond the boundaries of the cluster itself.
In the shocked medium inside the bubble, strong magnetohydrodynamic (MHD) turbulences provide the conditions for particle acceleration via the second-order Fermi mechanism \citep[e.g.,][]{Bykov2014,Vieu2022}.
With a maximum proton energy of \SI{200}{\TeV} and a source extent of $\sim$\SI{100}{\pc}, the Hillas criterion \citep{Hillas1984} implies for an average turbulent fluid velocity $u=\SI{100}{\km\per\second}$ a minimum magnetic field strength of $\sim$\SI{13}{\micro\gauss} in this scenario.
While the acceleration time scales are much longer compared to the case of acceleration at shocks inside the cluster, the process could -- under favourable circumstances -- generate CRs with up to PeV energies during several Myr, the typical life time of young clusters \citep{Bykov2020}.
For an adiabatically expanding wind, and a density of the ambient ISM $\rho_0$, the radius of the superbubble is given by $R_\mathrm{SB}=0.76\,(L_\mathrm{w}\,/\,\rho_0)^{1/5} \tau^{3/5}$ \citep{Weaver1977,Koo1992}, or $R_\mathrm{SB}\sim 256\,(\rho_0\,/\,1\,m_\mathrm{H}\,\si{\per\cubic\cm})^{-1/5}\si{\pc}$ for our assumptions (cf.\ Table~\ref{tab:wd1_pars}).
Adopting, for example, $\rho_0=5\,m_\mathrm{H}\,\si{\per\cubic\cm}$, we obtain $R_\mathrm{SB}\sim \SI{185}{\pc}$ -- a value more than two orders of magnitude larger than the half-mass radius of the cluster.
A superbubble with a size of this order around Westerlund~1 has not yet been revealed at other wavelengths.
Because its dimensions also exceed the extent of the $\gamma$-ray emission detected with H.E.S.S., an association of this emission with the entire superbubble seems disfavoured.
However, the assumption of a homogeneous medium is an oversimplification, and bubbles in a structured and possibly clumpy medium may have significantly different cooling rates and dynamics \citep[e.g.,][]{Chu2008}.
Moreover, more detailed models of superbubble evolution indicate that the simple estimate for their radius given above often over-predicts their true size \citep{Yadav2017,Vieu2022}.
If this is also the case for Westerlund~1, smaller-scale structures as for example the bubble-like feature `B3' in H~I data reported by \citet{Kothes2007} could be associated with the cluster.
Nevertheless, even in this case a connection between the superbubble and the $\gamma$-ray emission is not obvious (but conceivable, considering also that the $\gamma$-ray emission need not be uniform from the superbubble volume).

\subsubsection{Termination shock of collective cluster wind}
\label{sec:termination_shock}
Another possible site for the acceleration of CRs due to Westerlund~1, but outside the stellar cluster itself, is the termination shock of the collective cluster wind.
The termination shock forms where the pressure of the outgoing wind equals that of the ISM.
Recently, \citet{Morlino2021} have proposed that termination shocks of collective stellar cluster winds may be efficient sites of particle acceleration, and demonstrated that CRs with PeV energies could be produced in powerful clusters like Westerlund~1.
Considering again as above an adiabatic expansion, the radius of the termination shock is given by $R_\mathrm{TS}=0.92\,(L_\mathrm{w}\,/\,\rho_0)^{3/10} v_\mathrm{w}^{-1/2} \tau^{2/5}$ \citep{Koo1992}, or $R_\mathrm{TS}\sim 51\,(\rho_0\,/\,1\,m_\mathrm{H}\,\si{\per\cubic\cm})^{-3/10}\,\si{\pc}$ with our adopted parameter values.
Inserting $\rho_0=5\,m_\mathrm{H}\,\si{\per\cubic\cm}$ yields a radius of $R_\mathrm{TS}\sim \SI{32}{\pc}$.
Notably, this value coincides well with the radial distance of $\sim$$\SI{34}{\pc}$ at which we observe a peak in the $\gamma$-ray excess profiles (cf.\ Sect.~\ref{sec:maps_profiles} and Fig.~\ref{fig:radial_profiles}).
The scenario of particle acceleration at the cluster wind termination shock therefore provides a natural explanation for the shell-like structure exhibited by the $\gamma$-ray emission detected with H.E.S.S., and can furthermore reproduce its radial extent under reasonable assumptions.
The apparent asymmetry of the shell with respect to the position of Westerlund~1 could be caused, for example, by a gradient in the density of the surrounding ISM, or by a SN that occurred within the cluster towards the direction of the asymmetry.

Adopting the hadronic scenario, we find that the termination shock model is also viable in terms of the required energetics: with $L_\mathrm{w}\sim\SI{e39}{\erg\per\second}$ and $\tau\sim\SI{4}{\mega\year}$, the total available energy is $\sim$\SI{1.3e53}{\erg}, which in principle suffices to explain the required energy in CR protons, $W_p\sim \num{6e51}\,(n/\SI{1}{\per\cubic\cm})^{-1}\,\si{\erg}$ -- although, since the cooling time for protons exceeds the cluster lifetime, the acceleration process would need to be rather efficient, or the target density high.
Furthermore, the energe\-tics argument presupposes that the CRs can be confined within the $\gamma$-ray emission region over a significant fraction of the full cluster lifetime, which is not straightforward.
For instance, adopting Bohm diffusion, the diffusion length for protons is $L\sim \sqrt{6Dt}\sim 81\,(E/\SI{100}{\TeV})^{1/2}\,(B/\SI{10}{\micro\gauss})^{-1/2}\,(t/\SI{1}{\mega\year})^{1/2}\,\si{\pc}$ \citep{Chandrasekhar1943}, where we have neglected projection effects.
Hence, even for slow diffusion, to confine protons with energy \SI{200}{\TeV} -- our lower limit for the cut-off energy of the primary proton spectrum in a hadronic scenario -- within a region of radius $\sim$\SI{50}{\pc} for only \SI{1}{\mega\year} already requires a rather large magnetic field strength of $B\sim \SI{50}{\micro\gauss}$.
Nevertheless, taking into account, for example, an additional smearing due to the transformation from protons to $\gamma$ rays, the observed $\gamma$-ray morphology can be reproduced with a not-too-extreme assumption on the magnetic field, provided that the protons do not diffuse too fast.
However, as already mentioned in Sect.~\ref{sec:wind_wind}, the hadronic scenario is challenged further by the absence of a correlation of the $\gamma$-ray emission with the gas observed in the region.

Interestingly, the finding that the expected location of the termination shock coincides with the shell-like structure of the measured $\gamma$-ray emission also renders possible an interpretation within the leptonic scenario.
This is because the geometry of the acceleration site naturally explains the relatively large extent of the emission region and its rather complex structure, which are otherwise not easy to accommodate in a leptonic scenario.
The scenario is also feasible energetically; even in the presence of a \SI{10}{\micro\gauss} magnetic field, the required power of $L_e\sim \SI{1.7e36}{\erg\per\second}$ (cf.\ Sect.~\ref{sec:spectra}) is easily provided by the cluster wind if the acceleration efficiency for electrons is of order 0.1\%.
However, as the accelerated electrons emit synchrotron radiation, the scenario is subject to constraints from observations at the corresponding wavelengths.
For example, from measurements by the \textit{Planck} satellite at a frequency of \SI{30}{\giga\hertz}\footnote{We have used the full-sky frequency map at \SI{30}{\giga\hertz} available through the Planck Legacy Archive, \url{http://pla.esac.esa.int/pla}.}, at which the radiation is dominated by synchrotron emission \citep{Planck_LFI_2018}, we infer an average intensity within $1^\circ$ around Westerlund~1 of \SI{0.55}{\mega\Jy\per\steradian}.
For a magnetic field of \SI{10}{\micro\gauss}, electrons with energies around \SI{0.01}{\TeV} emit synchrotron radiation at \SI{30}{\giga\hertz}.
Assuming that the electron spectrum extends down to these energies, the predicted intensity at \SI{30}{\giga\hertz} is $\sim \SI{0.3}{\mega\Jy\per\steradian}$.
Considering furthermore that only part of the emission detected with \textit{Planck} originates from the vicinity of Westerlund~1, we conclude that the \textit{Planck} measurements imply either a magnetic field strength smaller than \SI{10}{\micro\gauss} or a low-energy cut-off of the primary electron spectrum.
Finally, it is worth noting in this regard that for another superbubble detected at TeV energies, 30~Dor~C in the Large Magellanic Cloud, a synchrotron shell has been detected using X-ray measurements, and a leptonic scenario was found to be favoured to explain the TeV $\gamma$-ray emission \citep[][and references therein]{Kavanagh2019}.

\subsection{Escape of particles from the emission region}
So far we have assumed that particles accelerated in or around Westerlund~1 are confined over the lifetime of the cluster.
If a significant fraction of accelerated particles can escape there are several important consequences:
\begin{enumerate}[(i)]
  \item $\gamma$-ray emission would be expected outside the system, in the case of hadronic CRs in particular in molecular clouds;
  \item in the (most likely) case of energy-dependent escape, the inferred spectrum of particles within the system would be softened with respect to the injection spectrum by the energy-dependent escape probability;
  \item the total energy requirements would be increased.
\end{enumerate}
There is little evidence for (i) in the maps shown in Fig.~\ref{fig:flux_maps} -- with the possible exception of the emission region east of region `C', which, however, does not coincide with a molecular cloud at the nominal distance of Westerlund~1 (cf.\ Fig.~\ref{fig:hi_co_map_dist_3d9}).
The inferred injection indices for electrons and protons of $\Gamma_e\sim$3.0\footnote{
  The electron spectral index derived in the \textsc{naima} fit corresponds to the present-time population of electrons, whose energy spectrum is steepened with respect to the injection spectrum due to cooling.
}
and $\Gamma_p\sim$2.3 seem broadly consistent with acceleration theory \citep[e.g.,][]{Bell2013}, so there seems to be no indication for (ii), although a modest variation of the spectrum is tolerable within the precision of the H.E.S.S.\ measurement.
Finally, as the total energy requirement is already a challenge in most acceleration scenarios under the assumption of confinement, there also seems to be little room for (iii).
Thus, while not entirely inconceivable, we at least find no indications for particles escaping from the emission region.

In the absence of evidence for particle escape we need to consider if confinement over the cluster lifetime is reasonable or not.
As already discussed in Sect.~\ref{sec:termination_shock}, this is not straightforward in the case of CR protons, which in a disordered magnetic field would diffuse much too quickly.
A possible way to circumvent this problem would be a magnetic field topology in which field lines are preferentially in the plane orthogonal to the radial direction, which can substantially inhibit the radial diffusion.

\section{Conclusion}
\label{sec:conclusion}

We have presented a detailed analysis of HESS~J1646$-$458, a very-high-energy $\gamma$-ray source positionally coincident with the young massive stellar cluster Westerlund~1.
HESS~J1646$-$458 is largely extended ($\sim$$2^\circ$ in diameter), and exhibits a complex, shell-like structure, with Westerlund~1 close to its centre.
We found no indications for energy-dependent morphology.
The energy spectrum of HESS~J1646$-$458 extends to at least several tens of TeV, with a spectral index of $\sim$$-2.3$, and a gradual steepening above $\sim$\SI{10}{\TeV}.
Energy spectra extracted within 16 signal regions across the source region are very similar to each other, reinforcing the observation that the morphology of HESS~J1646$-$458 does not vary with $\gamma$-ray energy.

In a hadronic scenario with CR protons producing the $\gamma$ rays, the observed $\gamma$-ray spectrum implies proton energies in excess of several hundred TeV.
However, our analysis of the H~I and CO emission around Westerlund~1 indicates no clear correlation between hydrogen gas clouds and the $\gamma$-ray emission, as would be expected to some degree within such a scenario.
Nevertheless, in particular due to uncertainties related to the distribution of target gas, a hadronic scenario remains viable in principle.
On the other hand, the lack of significant energy-dependent morphology of the $\gamma$-ray emission represents a challenge for an interpretation within an IC-dominated, leptonic scenario.

Investigating the possible physical counterparts to HESS~J1646$-$458, we found that -- while the energetic pulsars PSR~J1648$-$4611 and PSR~J1650$-$4601 may be contributing to the emission in their immediate surroundings -- no other known object besides Westerlund~1 can be made responsible for the bulk of the $\gamma$-ray emission.
Particle acceleration due to the cluster may occur at various possible sites: at wind-wind or SN-wind interaction shocks within the cluster, at turbulences in the superbubble excavated by the collective cluster wind, or at the termination shock of the cluster wind.
Models in which the acceleration takes place within the cluster generally need to overcome the problem of transporting the accelerated CRs into the larger region from which we observe the $\gamma$-ray emission without too severe energy losses, and explain the fact that the $\gamma$-ray emission does not peak towards the cluster position -- in particular the latter argument rules out a leptonic scenario with continuous injection for this case.
Attributing the CR acceleration to the superbubble as a whole seems disfavoured because HESS~J1646$-$458 -- although largely extended -- is still significantly smaller than the expected size of the superbubble, which has, furthermore, so far eluded its detection at other wavelengths.
Therefore, we deem most attractive the scenario of CR acceleration at the cluster wind termination shock, because it provides a natural explanation for the shell-like morphology of HESS~J1646$-$458, and the wind is powerful enough to sustain the $\gamma$-ray emission.
Based on the available data, however, we are not able to firmly identify the acceleration mechanism at work.

Our results further support massive stellar clusters as CR accelerators, and motivate to investigate Westerlund~1 and other representatives of this class of objects more deeply in the future.
In particular, we encourage a deep and broad coverage in X-ray observations of the region around Westerlund~1, which may enable the identification of the cluster wind termination shock.
On the other hand, a more accurate measurement of the $\gamma$-ray emission in this region will be provided by the upcoming Cherenkov Telescope Array \citep{CTA2018}, which is designed to be an order of magnitude more sensitive than the H.E.S.S.\ experiment.
Exploiting the data from this and other observatories will be crucial in understanding the contribution of massive stellar clusters to the sea of CRs in the Milky Way.

\begin{acknowledgements}
The support of the Namibian authorities and of the University of Namibia in facilitating the construction and operation of H.E.S.S.\ is gratefully acknowledged, as is the support by
the German Ministry for Education and Research (BMBF),
the Max Planck Society,
the German Research Foundation (DFG),
the Helmholtz Association,
the Alexander von Humboldt Foundation,
the French Ministry of Higher Education, Research and Innovation,
the Centre National de la Recherche Scientifique (CNRS/IN2P3 and CNRS/INSU),
the Commissariat \`{a} l'\'{E}nergie atomique et aux \'{E}nergies alternatives (CEA),
the U.K.\ Science and Technology Facilities Council (STFC),
the Knut and Alice Wallenberg Foundation,
the Polish Ministry of Education and Science, agreement no.~2021/WK/06,
the South African Department of Science and Technology and National Research Foundation,
the University of Namibia,
the National Commission on Research, Science \& Technology of Namibia (NCRST),
the Austrian Federal Ministry of Education, Science and Research and the Austrian Science Fund (FWF),
the Australian Research Council (ARC),
the Japan Society for the Promotion of Science,
the University of Amsterdam, and
the Science Committee of Armenia grant 21AG-1C085.
We appreciate the excellent work of the technical support staff in Berlin, Zeuthen, Heidelberg, Palaiseau, Paris, Saclay, T\"{u}bingen and in Namibia in the construction and operation of the equipment. 
This work benefited from services provided by the H.E.S.S.\ Virtual Organisation, supported by the national resource providers of the EGI Federation.
This research made use of the \textsc{Gammapy}\footnote{\url{https://gammapy.org}} \citep{Deil2017,Deil2020}, \textsc{Astropy}\footnote{\url{https://www.astropy.org}} \citep{Robitaille2013,PriceWhelan2018}, \textsc{Matplotlib}\footnote{\url{https://matplotlib.org}} \citep{Hunter2007}, and \textsc{Naima}\footnote{\url{https://naima.readthedocs.io}} \citep{Zabalza2015} software packages.
\end{acknowledgements}

\bibliographystyle{aa}
\bibliography{westerlund}

\begin{thebibliography}{114}
\expandafter\ifx\csname natexlab\endcsname\relax\def\natexlab#1{#1}\fi

\bibitem[{Abdalla {et~al.}(2018{\natexlab{a}})Abdalla, Abramowski, Aharonian,
  {et~al.}}]{HESS_MassiveStars_2018}
Abdalla, H., Abramowski, A., Aharonian, F., {et~al.} 2018{\natexlab{a}}, \aap,
  612, A11

\bibitem[{Abdalla {et~al.}(2018{\natexlab{b}})Abdalla, Abramowski, Aharonian,
  {et~al.}}]{HESS_HGPS_2018}
Abdalla, H., Abramowski, A., Aharonian, F., {et~al.} 2018{\natexlab{b}}, \aap,
  612, A1

\bibitem[{Abdalla {et~al.}(2018{\natexlab{c}})Abdalla, Abramowski, Aharonian,
  {et~al.}}]{HESS_PWNpop_2018}
Abdalla, H., Abramowski, A., Aharonian, F., {et~al.} 2018{\natexlab{c}}, \aap,
  612, A2

\bibitem[{Abdalla {et~al.}(2020)Abdalla, Adam, Aharonian,
  {et~al.}}]{HESS_EtaCar_2020}
Abdalla, H., Adam, R., Aharonian, F., {et~al.} 2020, \aap, 635, A167

\bibitem[{Abdalla {et~al.}(2018{\natexlab{d}})Abdalla, Aharonian, Ait~Benkhali,
  {et~al.}}]{HESS_VelaPulsar_2018}
Abdalla, H., Aharonian, F., Ait~Benkhali, F., {et~al.} 2018{\natexlab{d}},
  \aap, 620, A66

\bibitem[{Abdalla {et~al.}(2019)Abdalla, Aharonian, Ait~Benkhali,
  {et~al.}}]{HESS_1825_2019}
Abdalla, H., Aharonian, F., Ait~Benkhali, F., {et~al.} 2019, \aap, 621, A116

\bibitem[{Abdo {et~al.}(2010)Abdo, Ackermann, Ajello, {et~al.}}]{FermiLAT2010}
Abdo, A.~A., Ackermann, M., Ajello, M., {et~al.} 2010, \apj, 723, 649

\bibitem[{Abdo {et~al.}(2007)Abdo, Allen, Berley, {et~al.}}]{Abdo2007}
Abdo, A.~A., Allen, B., Berley, D., {et~al.} 2007, \apjl, 658, L33

\bibitem[{Abdollahi {et~al.}(2020)Abdollahi, Acero, Ackermann,
  {et~al.}}]{FermiLAT2020}
Abdollahi, S., Acero, F., Ackermann, M., {et~al.} 2020, \apjs, 247, 33

\bibitem[{Abeysekara {et~al.}(2021)Abeysekara, Albert, Alfaro,
  {et~al.}}]{HAWC2021}
Abeysekara, A.~U., Albert, A., Alfaro, R., {et~al.} 2021, Nat. Astron., 5, 465

\bibitem[{Abeysekara {et~al.}(2018)Abeysekara, Archer, Aune,
  {et~al.}}]{VERITAS2018}
Abeysekara, A.~U., Archer, A., Aune, T., {et~al.} 2018, \apj, 861, 134

\bibitem[{Abramowski {et~al.}(2011)Abramowski, Acero, Aharonian,
  {et~al.}}]{HESS_Wd2_2011}
Abramowski, A., Acero, F., Aharonian, F., {et~al.} 2011, \aap, 525, A46

\bibitem[{Abramowski {et~al.}(2012)Abramowski, Acero, Aharonian,
  {et~al.}}]{HESS_Westerlund1_2012}
Abramowski, A., Acero, F., Aharonian, F., {et~al.} 2012, \aap, 537, A114

\bibitem[{Abramowski {et~al.}(2014{\natexlab{a}})Abramowski, Aharonian,
  Ait~Benkhali, {et~al.}}]{HESS_Diffuse_2014}
Abramowski, A., Aharonian, F., Ait~Benkhali, F., {et~al.} 2014{\natexlab{a}},
  \prd, 90, 122007

\bibitem[{Abramowski {et~al.}(2014{\natexlab{b}})Abramowski, Aharonian,
  Ait~Benkhali, {et~al.}}]{HESS_1641_2014}
Abramowski, A., Aharonian, F., Ait~Benkhali, F., {et~al.} 2014{\natexlab{b}},
  \apjl, 794, L1

\bibitem[{Abramowski {et~al.}(2014{\natexlab{c}})Abramowski, Aharonian,
  Ait~Benkhali, {et~al.}}]{HESS_1640_2014}
Abramowski, A., Aharonian, F., Ait~Benkhali, F., {et~al.} 2014{\natexlab{c}},
  \mnras, 439, 2828

\bibitem[{Abramowski {et~al.}(2015)Abramowski, Aharonian, Ait~Benkhali,
  {et~al.}}]{HESS_30DorC_2015}
Abramowski, A., Aharonian, F., Ait~Benkhali, F., {et~al.} 2015, Science, 347,
  406

\bibitem[{Acharya {et~al.}(2018)Acharya, Agudo, Al~Samarai, {et~al.}}]{CTA2018}
Acharya, B.~S., Agudo, I., Al~Samarai, I., {et~al.} 2018, {Science with the
  Cherenkov Telescope Array} (World Scientific Publishing)

\bibitem[{Ackermann {et~al.}(2011)Ackermann, Ajello, Allafort,
  {et~al.}}]{FermiLAT2011}
Ackermann, M., Ajello, M., Allafort, A., {et~al.} 2011, Science, 334, 1103

\bibitem[{Ackermann {et~al.}(2012)Ackermann, Ajello, Atwood,
  {et~al.}}]{FermiLAT2012}
Ackermann, M., Ajello, M., Atwood, W.~B., {et~al.} 2012, \apj, 750, 3

\bibitem[{Aghakhanloo {et~al.}(2020)Aghakhanloo, Murphy, Smith, Parejko,
  Diaz-Rodriguez, Drout, Groh, Guzman, \& Stassun}]{Aghakhanloo2020}
Aghakhanloo, M., Murphy, J.~W., Smith, N., {et~al.} 2020, \mnras, 492, 2497

\bibitem[{Aghakhanloo {et~al.}(2021)Aghakhanloo, Murphy, Smith, Parejko,
  D\'{i}az-Rodr\'{i}guez, Drout, Groh, Guzman, \& Stassun}]{Aghakhanloo2021}
Aghakhanloo, M., Murphy, J.~W., Smith, N., {et~al.} 2021, Res. Notes ASS, 5, 14

\bibitem[{Aharonian {et~al.}(2006)Aharonian, Akhperjanian, Bazer-Bachi,
  {et~al.}}]{HESS_Crab_2006}
Aharonian, F., Akhperjanian, A.~G., Bazer-Bachi, A.~R., {et~al.} 2006, \aap,
  457, 899

\bibitem[{Aharonian {et~al.}(2007)Aharonian, Akhperjanian, Bazer-Bachi,
  {et~al.}}]{HESS_Wd2_2007}
Aharonian, F., Akhperjanian, A.~G., Bazer-Bachi, A.~R., {et~al.} 2007, \aap,
  467, 1075

\bibitem[{Aharonian {et~al.}(2019)Aharonian, Yang, \& de~O\~{n}a
  Wilhelmi}]{Aharonian2019}
Aharonian, F., Yang, R., \& de~O\~{n}a Wilhelmi, E. 2019, Nature Astronomy, 3,
  561

\bibitem[{Akrami {et~al.}(2020)Akrami, Arg\"{u}eso, Ashdown,
  {et~al.}}]{Planck_LFI_2018}
Akrami, Y., Arg\"{u}eso, F., Ashdown, M., {et~al.} 2020, \aap, 641, A2

\bibitem[{Amenomori {et~al.}(2021)Amenomori, Bao, Bi, {et~al.}}]{Tibet2021}
Amenomori, M., Bao, Y.~W., Bi, X.~J., {et~al.} 2021, \prl, 126, 141101

\bibitem[{Ballet {et~al.}(2020)Ballet, Burnett, Digel, \& Lott}]{FermiLAT2020a}
Ballet, J., Burnett, T.~H., Digel, S.~W., \& Lott, B. 2020, {Fermi Large Area
  Telescope Fourth Source Catalog Data Release 2},
  \href{https://arxiv.org/abs/2005.11208}{arXiv:2005.11208}

\bibitem[{Beasor {et~al.}(2021)Beasor, Davies, Smith, Gehrz, \&
  Figer}]{Beasor2021}
Beasor, E.~R., Davies, B., Smith, N., Gehrz, R., \& Figer, D. 2021, \apj, 912,
  16

\bibitem[{Belczynski \& Taam(2008)}]{Belczynski2008}
Belczynski, K. \& Taam, R.~E. 2008, \apj, 685, 400

\bibitem[{Bell(2013)}]{Bell2013}
Bell, A.~R. 2013, Astropart.\ Phys., 43, 56

\bibitem[{Bhadra {et~al.}(2022)Bhadra, Gupta, Nath, \& Sharma}]{Bhadra2022}
Bhadra, S., Gupta, S., Nath, B.~B., \& Sharma, P. 2022, \mnras, 510, 5579

\bibitem[{Bolatto {et~al.}(2013)Bolatto, Wolfire, \& Leroy}]{Bolatto2013}
Bolatto, A.~D., Wolfire, M., \& Leroy, A.~K. 2013, \araa, 51, 207

\bibitem[{Braiding {et~al.}(2018)Braiding, Wong, Maxted, Romano, Burton,
  Blackwell, Filipovi\'c, Freeman, Indermuehle, Lau, Rebolledo, Rowell,
  Snoswell, Tothill, Voisin, \& de~Wilt}]{Braiding2018}
Braiding, C., Wong, G.~F., Maxted, N.~I., {et~al.} 2018, \pasa, 35, e029

\bibitem[{Brandner {et~al.}(2008)Brandner, Clark, Stolte, Waters, Negueruela,
  \& Goodwin}]{Brandner2008}
Brandner, W., Clark, J.~S., Stolte, A., {et~al.} 2008, \aap, 478, 137

\bibitem[{Breuhaus {et~al.}(2022)Breuhaus, Hinton, Joshi, Reville, \&
  Schoorlemmer}]{Breuhaus2022}
Breuhaus, M., Hinton, J.~A., Joshi, V., Reville, B., \& Schoorlemmer, H. 2022,
  \aap, 661, A72

\bibitem[{Bykov(2014)}]{Bykov2014}
Bykov, A.~M. 2014, \aapr, 22, 77

\bibitem[{Bykov {et~al.}(2015)Bykov, Ellison, Gladilin, \& Osipov}]{Bykov2015}
Bykov, A.~M., Ellison, D.~C., Gladilin, P.~E., \& Osipov, S.~M. 2015, \mnras,
  453, 113

\bibitem[{Bykov {et~al.}(2020)Bykov, Marcowith, Amato, Kalyashova, Kruijssen,
  \& Waxman}]{Bykov2020}
Bykov, A.~M., Marcowith, A., Amato, E., {et~al.} 2020, Space Sci. Rev., 216, 42

\bibitem[{Cao {et~al.}(2021)Cao, Aharonian, An, {et~al.}}]{LHAASO2021}
Cao, Z., Aharonian, F.~A., An, Q., {et~al.} 2021, Nature, 594, 33

\bibitem[{Cesarsky \& Montmerle(1983)}]{Cesarsky1983}
Cesarsky, C.~J. \& Montmerle, T. 1983, Space Sci. Rev., 36, 173

\bibitem[{Chandrasekhar(1943)}]{Chandrasekhar1943}
Chandrasekhar, S. 1943, Rev. Mod. Phys., 15, 1

\bibitem[{{Chu}(2008)}]{Chu2008}
{Chu}, Y.-H. 2008, in Proc. Int. Astron. Union, ed. F.~{Bresolin}, P.~A.
  {Crowther}, \& J.~{Puls}, Vol. 250, 341--354

\bibitem[{Clark {et~al.}(2008)Clark, Muno, Negueruela, Dougherty, Crowther,
  Goodwin, \& de~Grijs}]{Clark2008}
Clark, J.~S., Muno, M.~P., Negueruela, I., {et~al.} 2008, \aap, 477, 147

\bibitem[{Clark {et~al.}(2019)Clark, Najarro, Negueruela, Ritchie,
  Gonz\'{a}lez-Fern\'{a}ndez, \& Lohr}]{Clark2019}
Clark, J.~S., Najarro, F., Negueruela, I., {et~al.} 2019, \aap, 623, A83

\bibitem[{Clark {et~al.}(2005)Clark, Negueruela, Crowther, \&
  Goodwin}]{Clark2005}
Clark, J.~S., Negueruela, I., Crowther, P., \& Goodwin, S.~P. 2005, \aap, 434,
  949

\bibitem[{Crowther {et~al.}(2006)Crowther, Hadfield, Clark, Negueruela, \&
  Vacca}]{Crowther2006}
Crowther, P.~A., Hadfield, L.~J., Clark, J.~S., Negueruela, I., \& Vacca, W.~D.
  2006, \mnras, 372, 1407

\bibitem[{Dame {et~al.}(2001)Dame, Hartmann, \& Thaddeus}]{Dame2001}
Dame, T.~M., Hartmann, D., \& Thaddeus, P. 2001, \apj, 547, 792

\bibitem[{Davies \& Beasor(2019)}]{Davies2019}
Davies, B. \& Beasor, E.~R. 2019, \mnras, 486, L10

\bibitem[{de~Naurois \& Rolland(2009)}]{deNaurois2009}
de~Naurois, M. \& Rolland, L. 2009, Astropart.\ Phys., 32, 231

\bibitem[{Deil {et~al.}(2020)Deil, Donath, Terrier, {et~al.}}]{Deil2020}
Deil, C., Donath, A., Terrier, R., {et~al.} 2020, {gammapy/gammapy: v0.17},
  {Zenodo. \url{https://doi.org/10.5281/zenodo.4701492}}

\bibitem[{Deil {et~al.}(2018)Deil, Wood, Hassan, Boisson, Contreras,
  Kn\"odlseder, Khelifi, King, \& Mohrmann}]{Deil2018}
Deil, C., Wood, M., Hassan, T., {et~al.} 2018, {Data formats for gamma-ray
  astronomy - version 2.0}, {Zenodo.
  \url{https://doi.org/10.5281/zenodo.1409831}}

\bibitem[{Deil {et~al.}(2017)Deil, Zanin, Lefaucheur, {et~al.}}]{Deil2017}
Deil, C., Zanin, R., Lefaucheur, J., {et~al.} 2017, in Proc. 35th Int. Cosmic
  Ray Conf. (ICRC2017), 766

\bibitem[{Eichler \& Usov(1993)}]{Eichler1993}
Eichler, D. \& Usov, V. 1993, \apj, 402, 271

\bibitem[{Forman {et~al.}(1978)Forman, Jones, Cominsky, Julien, Murray, Peters,
  Tananbaum, \& Giacconi}]{Forman1978}
Forman, W., Jones, C., Cominsky, L., {et~al.} 1978, \apjs, 38, 357

\bibitem[{Fujii \& Portegies~Zwart(2016)}]{Fujii2016}
Fujii, M.~S. \& Portegies~Zwart, S. 2016, \apj, 817, 4

\bibitem[{Gupta {et~al.}(2018)Gupta, Nath, Sharma, \& Eichler}]{Gupta2018}
Gupta, S., Nath, B.~B., Sharma, P., \& Eichler, D. 2018, \mnras, 473, 1537

\bibitem[{Heyer \& Dame(2015)}]{Heyer2015}
Heyer, M. \& Dame, T.~M. 2015, \araa, 53, 583

\bibitem[{Hillas(1984)}]{Hillas1984}
Hillas, A.~M. 1984, \araa, 22, 425

\bibitem[{Hinton \& Hofmann(2009)}]{Hinton2009}
Hinton, J.~A. \& Hofmann, W. 2009, \araa, 47, 523

\bibitem[{Holler {et~al.}(2015)Holler, Berge, van Eldik, {et~al.}}]{Holler2015}
Holler, M., Berge, D., van Eldik, C., {et~al.} 2015, in Proc. 34th Int. Cosmic
  Ray Conf. (ICRC2015), 847

\bibitem[{Hunter(2007)}]{Hunter2007}
Hunter, J.~D. 2007, Comput. Sci. Eng., 9, 90

\bibitem[{Israel {et~al.}(2007)Israel, Campana, Dall'Osso, Muno, Cummings,
  Perna, \& Stella}]{Israel2007}
Israel, G.~L., Campana, S., Dall'Osso, S., {et~al.} 2007, \apj, 664, 448

\bibitem[{Kantzas {et~al.}(2022)Kantzas, Markoff, Lucchini, Ceccobello,
  Grinberg, Connors, \& Uttley}]{Kantzas2022}
Kantzas, D., Markoff, S., Lucchini, M., {et~al.} 2022, \mnras, 510, 5187

\bibitem[{Kavanagh {et~al.}(2011)Kavanagh, Norci, \& Meurs}]{Kavanagh2011}
Kavanagh, P.~J., Norci, L., \& Meurs, E. J.~A. 2011, New Astron., 16, 461

\bibitem[{Kavanagh {et~al.}(2019)Kavanagh, Vink, Sasaki, Chu, Filipovi\'c, Ohm,
  Haberl, Manojlovic, \& Maggi}]{Kavanagh2019}
Kavanagh, P.~J., Vink, J., Sasaki, M., {et~al.} 2019, \aap, 621, A138

\bibitem[{Kissmann(2014)}]{Kissmann2014}
Kissmann, R. 2014, Astropart.\ Phys., 55, 37

\bibitem[{Kissmann(2022)}]{Kissmann2022}
Kissmann, R. 2022, private communication

\bibitem[{Kissmann {et~al.}(2017)Kissmann, Niederwanger, Reimer, \&
  Strong}]{Kissmann2017}
Kissmann, R., Niederwanger, F., Reimer, O., \& Strong, A.~W. 2017, AIP Conf.
  Proc., 1792, 070011

\bibitem[{Kissmann {et~al.}(2015)Kissmann, Werner, Reimer, \&
  Strong}]{Kissmann2015}
Kissmann, R., Werner, M., Reimer, O., \& Strong, A.~W. 2015, Astropart.\ Phys.,
  70, 39

\bibitem[{Kn\"{o}dlseder {et~al.}(2016)Kn\"{o}dlseder, Mayer, Deil, Cayrou,
  Owen, Kelley-Hoskins, Lu, Buehler, Forest, Louge, Siejkowski, Kosack, Gerard,
  Schulz, Martin, Sanchez, Ohm, Hassan, \& Brau-Nogu\'{e}}]{Knoedlseder2016}
Kn\"{o}dlseder, J., Mayer, M., Deil, C., {et~al.} 2016, \aap, 593, A1

\bibitem[{Koo \& McKee(1992)}]{Koo1992}
Koo, B.-C. \& McKee, C.~F. 1992, \apj, 388, 93

\bibitem[{Kothes \& Dougherty(2007)}]{Kothes2007}
Kothes, R. \& Dougherty, S.~M. 2007, \aap, 468, 993

\bibitem[{{Kramer} {et~al.}(2003){Kramer}, {Bell}, {Manchester}, {Lyne},
  {Camilo}, {Stairs}, {D'Amico}, {Kaspi}, {Hobbs}, {Morris}, {Crawford},
  {Possenti}, {Joshi}, {McLaughlin}, {Lorimer}, \& {Faulkner}}]{Kramer2003}
{Kramer}, M., {Bell}, J.~F., {Manchester}, R.~N., {et~al.} 2003, \mnras, 342,
  1299

\bibitem[{Li \& Ma(1983)}]{Li1983}
Li, T. \& Ma, Y. 1983, \apj, 272, 317

\bibitem[{Longair(1992)}]{Longair1992}
Longair, M.~S. 1992, {High energy astrophysics. Vol.1: Particles, photons and
  their detection} (Cambridge University Press)

\bibitem[{Manchester {et~al.}(2005)Manchester, Hobbs, Teoh, \&
  Hobbs}]{Manchester2005}
Manchester, R.~N., Hobbs, G.~B., Teoh, A., \& Hobbs, M. 2005, \aj, 129, 1993

\bibitem[{Mart\'{i}-Devesa \& Reimer(2021)}]{MartiDevesa2021}
Mart\'{i}-Devesa, G. \& Reimer, O. 2021, \aap, 654, A44

\bibitem[{McClure-Griffiths {et~al.}(2005)McClure-Griffiths, Dickey, Gaensler,
  Green, Haverkorn, \& Strasser}]{McClureGriffiths2005}
McClure-Griffiths, N.~M., Dickey, J.~M., Gaensler, B.~M., {et~al.} 2005, \apjs,
  158, 178

\bibitem[{Mohrmann {et~al.}(2019)Mohrmann, Specovius, Tiziani, Funk, Malyshev,
  Nakashima, \& van Eldik}]{Mohrmann2019}
Mohrmann, L., Specovius, A., Tiziani, D., {et~al.} 2019, \aap, 632, A72

\bibitem[{Morlino {et~al.}(2021)Morlino, Blasi, Peretti, \&
  Cristofari}]{Morlino2021}
Morlino, G., Blasi, P., Peretti, E., \& Cristofari, P. 2021, \mnras, 504, 6096

\bibitem[{Muno {et~al.}(2006{\natexlab{a}})Muno, Clark, Crowther, Dougherty,
  de~Grijs, Law, McMillan, Morris, Negueruela, Pooley, Portegies~Zwart, \&
  Yusef-Zadeh}]{Muno2006a}
Muno, M.~P., Clark, J.~S., Crowther, P.~A., {et~al.} 2006{\natexlab{a}}, \apj,
  636, L41

\bibitem[{Muno {et~al.}(2006{\natexlab{b}})Muno, Law, Clark, Dougherty,
  de~Grijs, Portegies~Zwart, \& Yusef-Zadeh}]{Muno2006}
Muno, M.~P., Law, C., Clark, J.~S., {et~al.} 2006{\natexlab{b}}, \apj, 650, 203

\bibitem[{Negueruela {et~al.}(2022)Negueruela, Alfaro, Dorda, Marco,
  Apell\'{a}niz, \& Gonz\'{a}lez-Fern\'{a}ndez}]{Negueruela2022}
Negueruela, I., Alfaro, E.~J., Dorda, R., {et~al.} 2022, \aap, (accepted
  manuscript),
  doi:\href{https://doi.org/10.1051/0004-6361/202142985}{10.1051/0004-6361/202142985}

\bibitem[{Ohm {et~al.}(2013)Ohm, Hinton, \& White}]{Ohm2013}
Ohm, S., Hinton, J.~A., \& White, R. 2013, \mnras, 434, 2289

\bibitem[{Ohm {et~al.}(2009)Ohm, van Eldik, \& Egberts}]{Ohm2009}
Ohm, S., van Eldik, C., \& Egberts, K. 2009, Astropart.\ Phys., 31, 383

\bibitem[{Parizot {et~al.}(2004)Parizot, Marcowith, van~der Swaluw, Bykov, \&
  Tatischeff}]{Parizot2004}
Parizot, E., Marcowith, A., van~der Swaluw, E., Bykov, A.~M., \& Tatischeff, V.
  2004, \aap, 424, 747

\bibitem[{Parsons \& Hinton(2014)}]{Parsons2014}
Parsons, R.~D. \& Hinton, J.~A. 2014, Astropart.\ Phys., 56, 26

\bibitem[{Piron {et~al.}(2001)Piron, Djannati-Ata\"{i}, Punch, Tavernet,
  Barrau, Bazer-Bachi, Chounet, Debiais, Degrange, Dezalay, Espigat, Fabre,
  Fleury, Fontaine, Goret, Gouiffes, Khelifi, Malet, Masterson, Mohanty, Nuss,
  Renault, Rivoal, Rob, \& Vorobiov}]{Piron2001}
Piron, F., Djannati-Ata\"{i}, A., Punch, M., {et~al.} 2001, \aap, 374, 895

\bibitem[{Popescu {et~al.}(2017)Popescu, Yang, Tuffs, Natale, Rushton, \&
  Aharonian}]{Popescu2017}
Popescu, C.~C., Yang, R., Tuffs, R.~J., {et~al.} 2017, \mnras, 470, 2539

\bibitem[{Portegies~Zwart {et~al.}(2010)Portegies~Zwart, McMillan, \&
  Gieles}]{PortegiesZwart2010}
Portegies~Zwart, S.~F., McMillan, S. L.~W., \& Gieles, M. 2010, \araa, 48, 431

\bibitem[{{Price-Whelan} {et~al.}(2018){Price-Whelan}, {Sip{\H{o}}cz},
  {G{\"u}nther}, {et~al.}}]{PriceWhelan2018}
{Price-Whelan}, A.~M., {Sip{\H{o}}cz}, B.~M., {G{\"u}nther}, H.~M., {et~al.}
  2018, \aj, 156, 123

\bibitem[{Pshirkov(2016)}]{Pshirkov2016}
Pshirkov, M.~S. 2016, \mnras, 457, L99

\bibitem[{Rate {et~al.}(2020)Rate, Crowther, \& Parker}]{Rate2020}
Rate, G., Crowther, P.~A., \& Parker, R.~J. 2020, \mnras, 495, 1209

\bibitem[{Reimer {et~al.}(2006)Reimer, Pohl, \& Reimer}]{Reimer2006}
Reimer, A., Pohl, M., \& Reimer, O. 2006, \apj, 644, 1118

\bibitem[{{Robitaille} {et~al.}(2013){Robitaille}, {Tollerud}, {Greenfield},
  {et~al.}}]{Robitaille2013}
{Robitaille}, T.~P., {Tollerud}, E.~J., {Greenfield}, P., {et~al.} 2013, \aap,
  558, A33

\bibitem[{Rohlfs \& Wilson(2004)}]{Rohlfs2004}
Rohlfs, K. \& Wilson, T.~L. 2004, {Tools of Radio Astronomy}, 4th edn.
  ({Springer-Verlag Berlin Heidelberg New York})

\bibitem[{Sakai {et~al.}(2013)Sakai, Matsumoto, Haba, Kanou, \&
  Miyamoto}]{Sakai2013}
Sakai, M., Matsumoto, H., Haba, Y., Kanou, Y., \& Miyamoto, Y. 2013, Publ.
  Astron. Soc. Jpn., 65, 64

\bibitem[{Specovius(2021)}]{Specovius2021}
Specovius, A. 2021, {PhD~thesis}, Friedrich-Alexander-Universit\"at
  Erlangen-N\"urnberg

\bibitem[{Sun {et~al.}(2020{\natexlab{a}})Sun, Yang, Liang, Peng, Zhang, Wang,
  \& Aharonian}]{Sun2020}
Sun, X.-N., Yang, R.-Z., Liang, Y.-F., {et~al.} 2020{\natexlab{a}}, \aap, 639,
  A80

\bibitem[{Sun {et~al.}(2020{\natexlab{b}})Sun, Yang, \& Wang}]{Sun2020a}
Sun, X.-N., Yang, R.-Z., \& Wang, X.-Y. 2020{\natexlab{b}}, \mnras, 494, 3405

\bibitem[{Tavani {et~al.}(2009)Tavani, Sabatini, Pian, {et~al.}}]{AGILE2009}
Tavani, M., Sabatini, S., Pian, E., {et~al.} 2009, \apj, 698, L142

\bibitem[{Vieu {et~al.}(2022)Vieu, Gabici, Tatischeff, \&
  Ravikularaman}]{Vieu2022}
Vieu, T., Gabici, S., Tatischeff, V., \& Ravikularaman, S. 2022, \mnras, 512,
  1275

\bibitem[{Vink \& Gr\"afener(2012)}]{Vink2012}
Vink, J.~S. \& Gr\"afener, G. 2012, \apjl, 751, L34

\bibitem[{Weaver {et~al.}(1977)Weaver, McCray, Castor, Shapiro, \&
  Moore}]{Weaver1977}
Weaver, R., McCray, R., Castor, J., Shapiro, P., \& Moore, R. 1977, \apj, 218,
  377

\bibitem[{Werner {et~al.}(2013)Werner, Reimer, Reimer, \& Egberts}]{Werner2013}
Werner, M., Reimer, O., Reimer, A., \& Egberts, K. 2013, \aap, 555, A102

\bibitem[{Westerlund(1961)}]{Westerlund1961}
Westerlund, B. 1961, Publ. Astron. Soc. Pac., 73, 51

\bibitem[{Wolfire {et~al.}(2010)Wolfire, Hollenbach, \& McKee}]{Wolfire2010}
Wolfire, M.~G., Hollenbach, D., \& McKee, C.~F. 2010, \apj, 716, 1191

\bibitem[{Yadav {et~al.}(2017)Yadav, Mukherjee, Sharma, \& Nath}]{Yadav2017}
Yadav, N., Mukherjee, D., Sharma, P., \& Nath, B.~B. 2017, \mnras, 465, 1720

\bibitem[{Yang {et~al.}(2018)Yang, de~O\~{n}a Wilhelmi, \&
  Aharonian}]{Yang2018}
Yang, R., de~O\~{n}a Wilhelmi, E., \& Aharonian, F. 2018, \aap, 611, A77

\bibitem[{Yang \& Aharonian(2017)}]{Yang2017}
Yang, R.-Z. \& Aharonian, F. 2017, \aap, 600, A107

\bibitem[{Yang \& Wang(2020)}]{Yang2020}
Yang, R.-Z. \& Wang, Y. 2020, \aap, 640, A60

\bibitem[{Zabalza(2015)}]{Zabalza2015}
Zabalza, V. 2015, in Proc. 34th Int. Cosmic Ray Conf. (ICRC2015), 922

\bibitem[{Zorn(2019)}]{Zorn2019}
Zorn, J. 2019, {PhD thesis}, Ruprecht-Karls-Universität Heidelberg

\end{thebibliography}

\begin{appendix}

\section[Contribution from Galactic diffuse gamma-ray emission]{Contribution from Galactic diffuse $\gamma$-ray emission}
\label{sec:appendix_picard}

We describe in this appendix our estimation for the contribution of Galactic diffuse emission to the $\gamma$-ray emission from HESS~J1646$-$458, based on a prediction obtained with the \textsc{Picard} CR propagation code \citep{Kissmann2014}.
The \textsc{Picard} simulation is based on an analytical, continuous distribution of CR sources that follows the spiral arms of the Milky Way \citep[for more details, see][]{Kissmann2015}.
\citet{Kissmann2017} used these simulations to derive predictions for the flux of diffuse $\gamma$ rays from three different processes: bremsstrahlung and IC emission from CR electrons, and the decay of neutral pions (and other short-lived particles) created in interactions of hadronic CRs with the interstellar medium.
Template maps of these predictions were provided to us by \citet{Kissmann2022}.
The predictions depend on various model assumptions (e.g., the CR source distribution, the distribution of gas and radiation fields, \dots), meaning that in particular the predicted absolute flux levels are rather uncertain.
Nevertheless, we used the templates as an estimate for the level of diffuse emission in the Westerlund~1 region, noting that contributions larger than predicted would quickly lead to a worse agreement between our background model (which predicts the level of hadronic background, but does not take into account diffuse $\gamma$-ray emission) and the observed data (cf.\ Figs.~\ref{fig:sign_map_div} and \ref{fig:sign_dist}).

\begin{figure}[ht]
  \centering
  \includegraphics{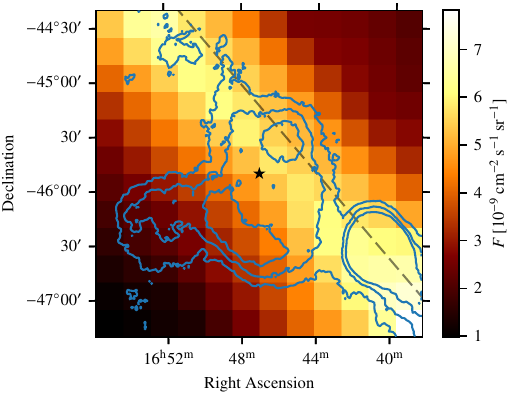}
  \caption{
    Prediction of total diffuse $\gamma$-ray flux above \SI{0.37}{\TeV} from the \textsc{Picard} code.
    The total emission includes contributions from bremsstrahlung, IC emission, and hadronic processes.
    The position of Westerlund~1 is marked by the black star symbol; the grey, dashed line shows the Galactic plane.
    Blue lines are contour lines of the flux map shown in Fig.~\ref{fig:flux_map}.
  }
  \label{fig:picard_flux_map}
\end{figure}

\begin{figure*}[pt]
  \centering
  \subfigure[Smoothing kernel: $0.22^\circ$ top hat]{
    \includegraphics{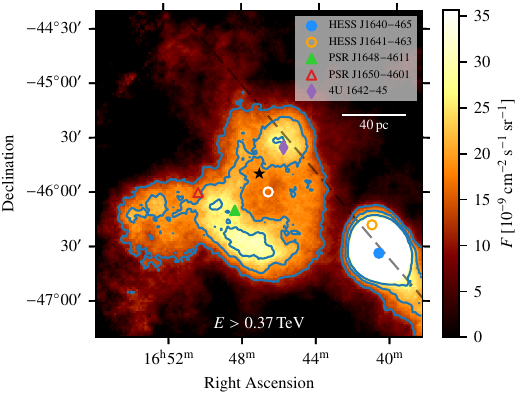}
    \label{fig:flux_map_picard_subtracted}
  }
  \subfigure[Smoothing kernel: $0.07^\circ$ Gaussian]{
    \includegraphics{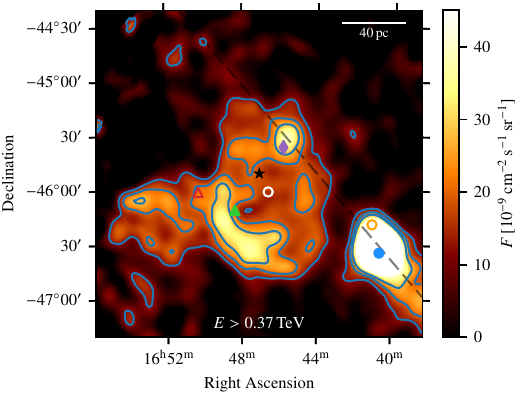}
    \label{fig:flux_map_hires_picard_subtracted}
  }\\
  \subfigure[Smoothing kernel: $0.22^\circ$ top hat]{
    \includegraphics{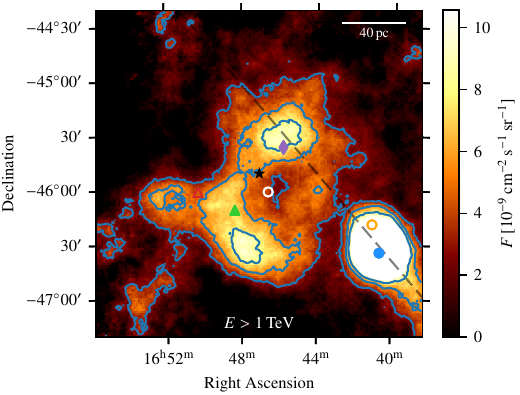}
    \label{fig:flux_map_1TeV_picard_subtracted}
  }
  \subfigure[Smoothing kernel: $0.22^\circ$ top hat]{
    \includegraphics{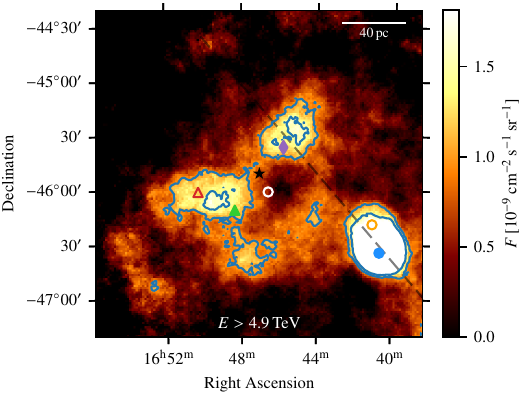}
    \label{fig:flux_map_5TeV_picard_subtracted}
  }
  \caption{
    Flux maps after subtraction of Galactic diffuse $\gamma$-ray emission.
    The maps are the same as in Fig.~\ref{fig:flux_maps}, except that the Galactic diffuse $\gamma$-ray flux as predicted by the \textsc{Picard} code has been subtracted.
    The colour axis scales and flux levels of contour lines are identical to those in Fig.~\ref{fig:flux_maps}.
  }
  \label{fig:flux_maps_picard_subtracted}
\end{figure*}

We show in Fig.~\ref{fig:picard_flux_map} the total predicted flux of diffuse $\gamma$ rays (i.e., the sum of the flux for each of the aforementioned processes), in the energy range above \SI{0.37}{\TeV}.
In this region and energy range, IC emission from CR electrons represents the dominant contribution to the Galactic diffuse emission.
In Fig.~\ref{fig:flux_maps_picard_subtracted}, we show the same flux maps as in Fig.~\ref{fig:flux_maps}, but with the diffuse $\gamma$-ray flux as predicted by the \textsc{Picard} code in the respective energy range subtracted.

For a more quantitative comparison, we integrated the $\gamma$-ray flux measured with H.E.S.S.\ and that predicted as diffuse emission by \textsc{Picard} in each of the 16 signal regions (a--p) defined in Fig.~\ref{fig:sign_map_boxes}, the resulting values are displayed in Table~\ref{tab:picard_boxes}.
Above the lowest threshold energy of \SI{0.37}{\TeV}, the ratio between the predicted diffuse emission and the measured $\gamma$-ray flux varies between $\sim$50\% for region `a' and $\sim$12\% for region `m'.
For higher energy thresholds we obtained smaller ratios in general, down to $\sim$4\% contamination above \SI{4.9}{\TeV} for region `j'.
Summing up the fluxes over all signal regions, we found for energy thresholds of \SI{0.37}{\TeV}, \SI{1}{\TeV}, and \SI{4.9}{\TeV} a ratio of $\sim$24\%, $\sim$17\%, and $\sim$8\%, respectively.
We stress again, however, that the absolute flux level of the diffuse emission is rather uncertain, and so are the estimates of its contribution to the total observed emission.

Finally, we note that, due to the omnipresence of the diffuse $\gamma$-ray emission along the Galactic plane, it is likely that at least part of it has been absorbed by the hadronic background model when we adjusted it to the observed data (cf.\ Sect.~\ref{sec:hess_data_analysis}).
This would reduce the measured $\gamma$-ray flux, and hence render the ratios $F_\mathrm{diff}/F_\mathrm{meas}$ listed in Table~\ref{tab:picard_boxes} overestimates, even if $F_\mathrm{diff}$ was precisely known.

\begin{table*}[pt]
  \centering
  \caption{Comparison of measured $\gamma$-ray flux and predicted diffuse flux for signal regions, above different energy thresholds.}
  \label{tab:picard_boxes}
  \begin{tabular}{cccccccccccc}
    \hline\hline
     & \multicolumn{3}{c}{$E > \SI{0.37}{\TeV}$} & & \multicolumn{3}{c}{$E > \SI{1}{\TeV}$} & & \multicolumn{3}{c}{$E > \SI{4.9}{\TeV}$}\\
    Signal region & $F_\mathrm{meas}$ & $F_\mathrm{diff}$ & $F_\mathrm{diff}/F_\mathrm{meas}$ & & $F_\mathrm{meas}$ & $F_\mathrm{diff}$ & $F_\mathrm{diff}/F_\mathrm{meas}$ & & $F_\mathrm{meas}$ & $F_\mathrm{diff}$ & $F_\mathrm{diff}/F_\mathrm{meas}$ \\\hline
    a & 34.8 & 17.5 & 0.50 && 7.32 & 3.12 & 0.43 && 0.457 & 0.218 & 0.48\\
    b & 38.5 & 18.1 & 0.47 && 7.64 & 3.24 & 0.42 && 0.546 & 0.227 & 0.42\\
    c & 64.8 & 17.6 & 0.27 && 19.3 & 3.14 & 0.16 && 1.86 & 0.219 & 0.12\\
    d & 56.3 & 15.6 & 0.28 && 21.2 & 2.76 & 0.13 && 3.61 & 0.191 & 0.05\\
    e & 26.1 & 12.5 & 0.48 && 10.6 & 2.19 & 0.21 && 1.38 & 0.150 & 0.11\\
    f & 48.3 & 13.7 & 0.28 && 9.00 & 2.44 & 0.27 && 2.33 & 0.170 & 0.07\\
    g & 75.3 & 15.9 & 0.21 && 19.4 & 2.83 & 0.15 && 2.69 & 0.197 & 0.07\\
    h & 71.8 & 17.0 & 0.24 && 18.4 & 3.03 & 0.17 && 2.44 & 0.211 & 0.09\\
    i & 60.0 & 8.39 & 0.14 && 13.8 & 1.47 & 0.11 && 2.04 & 0.101 & 0.05\\
    j & 84.4 & 10.9 & 0.13 && 20.3 & 1.91 & 0.09 && 3.74 & 0.132 & 0.04\\
    k & 80.3 & 13.8 & 0.17 && 20.1 & 2.44 & 0.12 && 2.72 & 0.170 & 0.06\\
    l & 74.1 & 18.3 & 0.25 && 16.8 & 3.27 & 0.19 && 2.72 & 0.228 & 0.08\\
    m & 50.4 & 6.27 & 0.12 && 7.36 & 1.08 & 0.15 && 1.24 & 0.074 & 0.06\\
    n & 38.6 & 8.21 & 0.21 && 8.50 & 1.43 & 0.17 && 1.75 & 0.098 & 0.06\\
    o & 72.9 & 11.0 & 0.15 && 20.2 & 1.94 & 0.10 && 2.73 & 0.135 & 0.05\\
    p & 58.4 & 15.8 & 0.27 && 14.4 & 2.81 & 0.20 && 2.04 & 0.196 & 0.10\\
    \hline
    All & 935 & 221 & 0.24 && 234 & 39.1 & 0.17 && 34.3 & 2.72 & 0.08\\
    \hline
  \end{tabular}
  \tablefoot{
    See Fig.~\ref{fig:sign_map_boxes} for the definition of the signal regions.
    $F_\mathrm{meas}$ and $F_\mathrm{diff}$ denote the measured and predicted diffuse $\gamma$-ray flux, respectively, in units of \SI{e-11}{\per\square\cm\per\second}.
  }
\end{table*}

\clearpage

\section{Radio maps for other velocity intervals}
\label{sec:appendix_radio}

\begin{figure}[ht]
  \centering
  \includegraphics{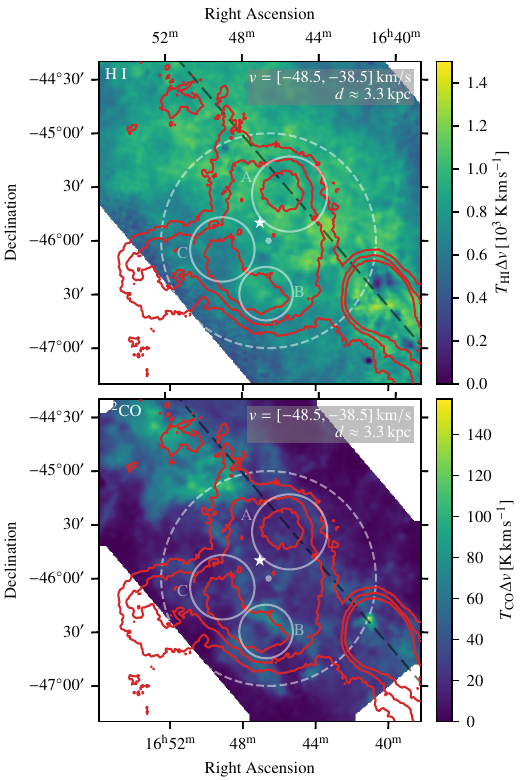}
  \caption{Same as Fig.~\ref{fig:hi_co_map_dist_3d9}, but adopting an interval in velocity with respect to the local standard of rest of $v=[-48.5, -38.5]\,\si{\km\per\second}$.}
  \label{fig:hi_co_map_dist_3d3}
\end{figure}

\begin{figure}[ht]
  \centering
  \includegraphics{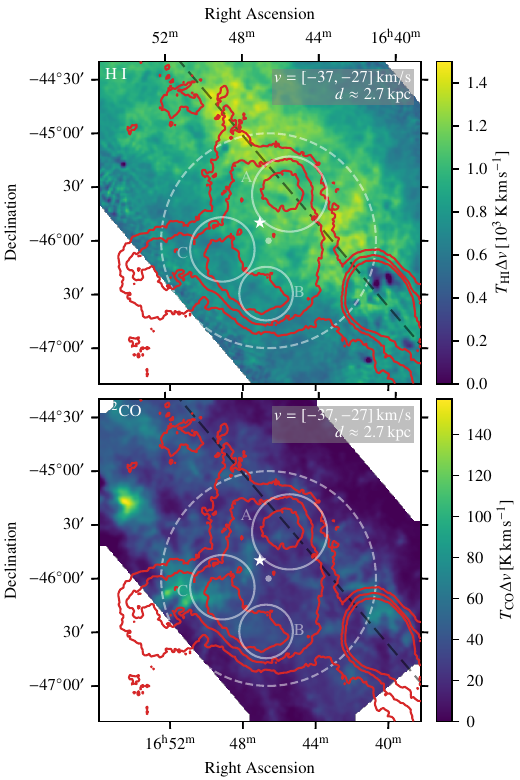}
  \caption{Same as Fig.~\ref{fig:hi_co_map_dist_3d9}, but adopting an interval in velocity with respect to the local standard of rest of $v=[-37, -27]\,\si{\km\per\second}$.}
  \label{fig:hi_co_map_dist_2d7}
\end{figure}

As a comparison to Fig.~\ref{fig:hi_co_map_dist_3d9}, we show H~I and CO maps for velocity intervals of $v=[-48.5, -38.5]\,\si{\km\per\second}$ and $v=[-37, -27]\,\si{\km\per\second}$ in Figs.~\ref{fig:hi_co_map_dist_3d3} and~\ref{fig:hi_co_map_dist_2d7}, respectively.

\end{appendix}

\end{document}